\documentclass[a4paper,12pt]{article}

\pdfoutput=1
\usepackage[T1]{fontenc}
\usepackage{comment}
\usepackage{jcappub}
\usepackage{subcaption}
\usepackage{graphicx}
\usepackage{xcolor}
\usepackage{amssymb,amsmath,bm}
\usepackage[utf8]{inputenc}
\usepackage[export]{adjustbox}
\hypersetup{colorlinks=true, citecolor=blue}
\usepackage{multirow}
\usepackage{subcaption}
\usepackage{amsfonts}
\usepackage{siunitx}
\usepackage{tensor}
\usepackage{booktabs}
\usepackage{mathtools}
\usepackage{placeins}
\usepackage{anyfontsize}
\usepackage{float}
\usepackage{hyperref}
\usepackage{cleveref}
\usepackage{xspace}
\allowdisplaybreaks

\def\be{\begin{equation}}
\def\ee{\end{equation}}
\def\ba{\begin{eqnarray}}
\def\ea{\end{eqnarray}}

\newcommand\LCDM{$\Lambda$CDM\xspace}

\def\eg{{\frenchspacing\it e.g.}\xspace}

\newcommand{\CITE}[1]{[{\bf CITE}]}
\definecolor{darkviolet}{rgb}{0.58, 0.0, 0.83}

\usepackage[outercaption]{sidecap}
\usepackage{subfiles}

\usepackage[toc, page]{appendix}
\usepackage{tikz}
\usetikzlibrary{calc, intersections, through, arrows, backgrounds}
\usetikzlibrary{arrows.meta}

\hypersetup{
    colorlinks=true,
    citecolor=blue,
    linkcolor=blue,
    linktoc=page
}

\usepackage{graphicx}

\usepackage[normalem]{ulem}

\newcommand{\pb}{\texttt{PyBird}\xspace}

\title{\fontsize{20}{32}\selectfont{An efficient one-loop EFTofLSS framework for Vainshtein-screened Horndeski gravity \vspace{-0.2in} \\}}

\author[a]{\fontsize{15}{25}\selectfont Stan Verhoeve,}
\author[a]{Dani de Boe,}
\author[a,b]{Gen Ye,}
\author[a]{Alessandra Silvestri}

\affiliation[a]{Institute Lorentz, Leiden University, PO Box 9506, Leiden 2300 RA, The Netherlands}
\affiliation[b]{D\'{e}partement de Physique Th\'{e}orique, Universit\'{e} de Gen\`{e}ve,
24 quai Ernest-Ansermet, CH-1211 Gen\`{e}ve 4, Switzerland}
\emailAdd{\textcolor{blue}{sverhoeve@lorentz.leidenuniv.nl}}
\emailAdd{\textcolor{blue}{deboe@lorentz.leidenuniv.nl}}
\emailAdd{\textcolor{blue}{Gen.Ye@unige.ch}}
\emailAdd{\textcolor{blue}{silvestri@lorentz.leidenuniv.nl}}

\abstract{We present an extension of \texttt{PyBird} for one-loop large-scale structure
analyses of modified gravity models. We implement support for quasi-static,
Vainshtein-screened luminal Horndeski models (in EFTofDE and covariant
formalisms) and nDGP, and replace the Green's function approach with a direct
ODE method for computing the exact time-dependent functions entering the
perturbation kernels. The new implementation improves computational efficiency
while maintaining numerical consistency with the standard approach, and we
validate the resulting one-loop matter power spectrum against $N$-body
simulations for the $\alpha_i\propto\Omega_{\rm DE}$ parametrization. We apply
this framework to constrain the $\alpha_i \propto a^3$ and
$\alpha_i \propto \Omega_{\rm DE}$ parametrizations using Planck CMB, BOSS
full-shape, and DESI DR2 BAO data, finding that full-shape information
significantly tightens the constraints. We further showcase the pipeline for
the cubic Galileon and nDGP models, demonstrating its applicability beyond the
phenomenological amplitude parametrizations to covariant modified-gravity
theories. Finally, we assess the impact of the Einstein--de Sitter
approximation and find that exact time dependence can be retained at modest
computational cost, which may become relevant for future large-scale
structure surveys.}

\begin{document}

\maketitle
\section{Introduction}
\label{sec:introduction}
The current generation of Stage IV galaxy surveys, including Euclid~\cite{Euclid:2009,Euclid:2024yrr} and DESI~\cite{DESI:2016fyo}, is delivering increasingly precise measurements of the large-scale structure (LSS) of the Universe. While previous surveys primarily probed linear scales, these experiments significantly extend the observational reach into the  nonlinear regime. Fully exploiting their constraining power requires accurate theoretical predictions beyond linear perturbation theory, particularly on quasi-linear scales (\(k \lesssim 0.3\,h\,\mathrm{Mpc}^{-1}\)), and eventually on smaller scales where screening effects and baryonic physics become important. Focusing on the spectroscopic data, the full-shape galaxy power spectrum on quasi-linear scales can be accurately modeled using the Effective Field Theory of Large-Scale Structure (EFTofLSS)~\cite{Baumann:2010tm,Carrasco:2012cv,Senatore:2014eva,Perko:2016puo}. By systematically incorporating the effects of unresolved small-scale physics through effective parameters and counterterms, the EFTofLSS extends the validity of standard perturbation theory (SPT) and has emerged as one of the leading frameworks for the analysis of LSS data~\cite{Ivanov:2019pdj,DAmico:2019fhj,DAmico:2022gki,Cabass:2022wjy}.

In this paper, we apply the EFTofLSS framework to Horndeski gravity~\cite{Horndeski:1974wa}, a general Lagrangian for scalar-tensor theories with second order equations of motion that encompass most of the dynamical dark energy (DE) models which will undergo scrutiny with Stage IV surveys.
We focus on the subclass of luminal Horndeski models for which the speed of gravitational waves equals the speed of light ($c_T^2 = 1$), in agreement with the observational constraints from GW170817 and its electromagnetic counterpart~\cite{TheLIGOScientific:2017qsa,Baker:2017hug,LIGOScientific:2017zic,Goldstein:2017mmi}. Additionally, we assume Vainshtein screening~\cite{Vainshtein:1972sx} as the dominant screening mechanism  that restores GR on small scales.

Our theoretical predictions are obtained in two steps. First, we use $\mathcal{H}$--\texttt{EFTCAMB}~\cite{Hu:2013twa,Raveri:2014cka,Ye:2026qqf}, a public extension of \texttt{CAMB}~\cite{Lewis:1999bs}, to compute the linear matter power spectrum for any given luminal Horndeski model, either in its covariant or effective field theory of DE (EFTofDE) formulation. These results are then passed to a modified version of \texttt{PyBird}~\cite{DAmico:2020kxu}, which we extended to interface directly with $\mathcal{H}$--\texttt{EFTCAMB} and to accept user-defined functions describing the background and perturbation evolution. The resulting pipeline computes one-loop EFTofLSS predictions and galaxy-clustering observables. Our implementation is based on the framework developed in~\cite{Cusin_2018_nonlinear_eftofde,Cusin_2018_de_mg_lss} and follows a strategy similar to that of Ref.~\cite{Taule:2024bot}. However, our code is applicable to any quasi-static, Vainshtein-screened luminal Horndeski model with scale-independent growth, as well as to nDGP gravity~\cite{Dvali:2000hr,Koyama:2007ih}. In addition, we introduce a new approach in which time-dependent functions are computed directly, avoiding the intermediate construction of Green's functions used in the original \texttt{PyBird} implementation. This significantly improves computational efficiency while maintaining the accuracy of one-loop predictions at multiple redshifts. Furthermore, we evolve the perturbation-theory kernels using their exact time dependence rather than the Einstein--de Sitter approximation, whose accuracy may become model dependent in modified-gravity scenarios and relevant for future high-precision surveys.

Beyond theory predictions, \texttt{PyBird} includes a likelihood module based on BOSS galaxy-clustering data~\cite{BOSS:2015npt}. We adapt this likelihood to interface with our pipeline and perform parameter inference for the modified-gravity models considered here. Combining one-loop EFTofLSS predictions with BOSS measurements, a Planck baseline, and theoretical stability conditions, we derive new constraints on the parameter space of luminal Horndeski theories. Our analysis illustrates the constraining power of current LSS observations and establishes a framework for forthcoming Stage IV surveys, whose increased statistical precision will enable substantially more stringent tests of gravity on cosmological scales~\cite{Amendola:2016saw,Euclid:2019clj}.

The paper is structured as follows. In \cref{sec:EFTofDE}, we present the nonlinear EFTofDE framework. In \cref{sec:one_loop_spectrum}, we discuss the one-loop matter power spectrum in the EFTofLSS framework, introduce the Green’s function formalism, and present our novel direct computation of the time-dependent functions. In \cref{sec:numerical_implementation}, we describe the code, validation. model choices, and data sets. In \cref{sec:results}, we present a showcase of covariant models, derive constraints on EFTofDE models using Planck CMB, BOSS full-shape (FS), and BAO data, and investigate the impact of the EdS approximation compared to exact time evolution. Finally, \cref{sec:conclusion} contains the conclusions of this work.

\section{Non-linear evolution for extended theories}
\label{sec:EFTofDE}

Our goal is to develop a general framework for computing one-loop large-scale structure observables in any Vainshtein-screened luminal Horndeski model, irrespective of whether the theory is formulated through a covariant action or within the Effective Field Theory of Dark Energy (EFTofDE) formalism. For computational convenience, we implement the pipeline in the EFTofDE language, specifically using the time-dependent $\alpha$-basis~\cite{Bellini:2014fua}. The implementation is modular: the only required inputs are the relevant EFTofDE functions and background quantities, which can be supplied by any Einstein--Boltzmann solver or external module capable of computing them for the model under consideration. In this work, we obtain these quantities using $\mathcal{H}$--\texttt{EFTCAMB}, which can evolve both covariant Horndeski theories and models defined directly within the EFTofDE formalism.

In this section, we briefly review the main ingredients required to compute the matter power spectrum in modified gravity within the EFTofDE framework~\cite{Cusin_2018_de_mg_lss,Cusin_2018_nonlinear_eftofde}. Let us start from the matter sector, whose treatment remains unmodified. We will work in the Newtonian gauge for scalar perturbations
\begin{equation}
    \label{eq:sec2.metric_pert_newtonian_gauge}
    ds^2 = -(1+2\Phi)dt^2 + a^2(t)(1-2\Psi)\delta_{ij}dx^idx^j.
\end{equation}
and adopt the quasi-static approximation (QSA)~\cite{Sawicki:2015zya} and non-relativistic limit throughout. Using $\mathcal{H}$--\texttt{EFTCAMB} it is easy to check whether the QSA is a reasonable approximation for the model under study.  To construct the effective fluid equations, we follow the standard procedure, i.e. first we derive the full-field equations including both long- and short-wavelength perturbations, and then we smooth these fields to obtain equations for the long-wavelength components. This procedure introduces an effective stress tensor $\tau_s^{ij}$, which encapsulates the influence of small-scale physics on large-scale dynamics. The effective continuity and Euler equations of these smoothed fields are given as~\cite{Baumann:2010tm}
\begin{align}
    \label{eq:sec2.effective_continuity}
    &\dot{\delta} + \partial_i\left[(1+\delta)v^i\right]=0,
    \\
    \label{eq:sec2.effective_euler}
    &\dot{v}^i + \mathcal{H} v^i + v^j\partial_jv^i + \partial_i\Phi = -\frac{1}{\rho_m}\partial_j\tau_s^{ij},
\end{align}
where a dot denotes a derivative with respect to conformal time, $\delta\equiv\frac{\rho_m-\bar{\rho}_m}{\bar{\rho}_m}$ is the matter density contrast, and \(v^i\) the peculiar velocity field.

We now need to close the system of \cref{eq:sec2.effective_euler,eq:sec2.effective_continuity} with an equation relating the metric potential $\Phi$ to matter sources, namely the Poisson equation. The latter is derived from the Einstein equations, and therefore depends on the action for gravity. The EFTofDE action for  luminal Horndeski models reads
~\cite{Bloomfield:2013efa,Gleyzes:2013ooa,Cusin_2018_nonlinear_eftofde,Hu:2014oga}:
\begin{align}
\label{eq:sec2.horndeski_action_unitary}
\begin{aligned}
S = \int \mathrm{d}^4x \, \bigg\{ & \frac{m_0^2 (1+\Omega(\tau))}{2} \, {}^{(4)}R + \Lambda(\tau) - c(\tau)\, a^2 \delta g^{00} + \frac{m_0^2 H_0^2 \gamma_1(\tau)}{2} \, (a^2 \delta g^{00})^2 \\
& - \frac{m_0^2 H_0 \gamma_2(\tau)}{2} \, a^2 \delta K^{\mu}_{\ \mu} \delta g^{00} \bigg\} + S_m,
\end{aligned}
\end{align}
where we have adopted $\mathcal{H}$--\texttt{EFTCAMB} convention,  $\delta g^{00} \equiv 1 + g^{00}$, $K^\mu_\nu$ is the extrinsic curvature of constant time hypersurfaces, $\tau$ denotes conformal time, and $S_m$ is the matter action. We assume that dark matter is minimally coupled to gravity, ensuring the existence of a well-defined Jordan frame. The EFT functions can be expressed in dimensionless form as~\cite{Bellini:2014fua} \footnote{Note that our convention for $\alpha_B$ is $-\frac{1}{2}$ times the one of Ref.~\cite{Bellini:2014fua}.}
\begin{equation}
    \label{eq:sec2.alphas}
        \alpha_B  \equiv \frac{1}{2} \dfrac{a\gamma_2 H_0 + a\mathcal{H}\Omega^\prime}{\mathcal{H}(1+\Omega)}, \qquad \alpha_M \equiv \frac{a\Omega^\prime}{1+\Omega}, \qquad \alpha_K  \equiv \frac{2ca^2 + 4 m_0^2 H_0^2 \gamma_1 a^2}{m_0^2 (1+\Omega)\mathcal{H}^2},
\end{equation}
where $\mathcal{H}$ is the Hubble parameter in conformal time, and a prime denotes a derivative with respect to the scale factor $a$. These $\alpha$-functions capture the departures from \LCDM in the linear perturbation sector corresponding, respectively, to: kineticity $\alpha_K$ which encodes the independent dynamics of the scalar degree of freedom; braiding $\alpha_B$, indicating a coupling between the metric and the scalar degree of freedom; $\alpha_M$ signaling a running of the effective Planck mass.

The linear equations for scalar and tensor perturbations follow from the quadratic action obtained by expanding \cref{eq:sec2.horndeski_action_unitary} around the cosmological background. Beyond linear order, the nonlinear dynamics require the action to be expanded to higher order in the metric perturbations, as well as to cubic and quartic order in the EFT operators $\delta g^{00}$, $\delta K^\mu_{\ \nu}$, and $\delta R^\mu_{\ \nu}$. This derivation was performed in Refs.~\cite{Cusin_2018_de_mg_lss,Cusin_2018_nonlinear_eftofde} under the QSA  and within the gradient-expansion limits relevant for Vainshtein screening. For luminal Horndeski theories, no new operators need to be introduced:
the nonlinear interactions arise entirely from the higher-order expansion of the operators already appearing in \cref{eq:sec2.horndeski_action_unitary}. Since this action is written in unitary gauge, the first step is to transform it to the Newtonian gauge of \cref{eq:sec2.metric_pert_newtonian_gauge}, thereby making the scalar degree of freedom explicit. The scalar field can then be related to the matter density contrast by solving its nonlinear equation of motion, which displays characteristic Vainshtein terms.
The resulting nonlinear Poisson equation, truncated at $\mathcal{O}(\delta^3)$, takes the form~\cite{Cusin_2018_de_mg_lss,Cusin_2018_nonlinear_eftofde}
\begin{align}
    \label{eq:sec2.effective_poisson}
    \partial^2 \Phi = &\mathcal{H}^2 \Bigg\{
      \frac{3}{2} \Omega_m \mu_{\Phi} \delta + \left(\frac{3}{2} \Omega_m\right)^2 \mu_{\Phi,2} \left[ \delta^2 - (\partial^{-2} \partial_i \partial_j \delta)^2 \right] \nonumber \\
      &+ \left(\frac{3}{2} \Omega_m\right)^3 \mu_{\Phi,22} \left[ \delta - (\partial^{-2} \partial_i \partial_j \delta) \partial^{-2} \partial_i \partial_j \right] \left[ \delta^2 - (\partial^{-2} \partial_k \partial_l \delta)^2 \right] \nonumber \\
      &+ \left(\frac{3}{2} \Omega_m\right)^3\mu_{\Phi,3}\left[\delta^3 - 3\delta\left(\partial^{-2}\partial_i\partial_j\delta\right)^2 + 2\left(\partial^{-2}\partial_i\partial_j\delta\right)\left(\partial^{-2}\partial_k\partial_j\delta\right)\left(\partial^{-2}\partial_i\partial_k\delta\right)\right]\Bigg\} \nonumber \\
      &+ \mathcal{O}\left(\delta^4\right).
\end{align}
Here, an overbar denotes evaluation on the FLRW background, and the matter density parameter is given by
\begin{equation}
\Omega_m \equiv \frac{a^2 \bar{\rho}_m}{3m_0^2 (1+\Omega)\mathcal{H}^2}.
\end{equation}
The functions $\mu_{\Phi}$, $\mu_{\Phi,2}$, $\mu_{\Phi,22}$, and $\mu_{\Phi,3}$ encode modifications to the GR Poisson equation~\cite{Cusin_2018_nonlinear_eftofde,Cusin_2018_de_mg_lss}. For the luminal Horndeski subclass, these functions are given by
\begin{align}
\label{eq:mu-phi-eqs}
\mu_{\Phi}
&= 1 + \frac{2}{c_s^2 \alpha}\left(\alpha_M -\alpha_B \right)^2,
&\quad
\mu_{\Phi,2}
&= \frac{1}{2}(\alpha_M - 2\alpha_B)\left(\frac{2(\mu_{\Phi}-1)}{\alpha c_s^2}\right)^{3/2},
\nonumber\\
\mu_{\Phi,22}
&= \frac{4(\alpha_M - 2\alpha_B)^2 (\mu_{\Phi}-1)^2}{(\alpha c_s^2)^3},
&\quad
\mu_{\Phi,3}
&= 0\,,
\end{align}
where $c_s$ is the sound speed defined by
\begin{equation}
\frac{1}{2}\alpha c_s^2 \equiv
\alpha_M - \alpha_B(1-\alpha_M+\alpha_B)
- \frac{1}{\mathcal{H}^2}(1+\alpha_B)(\dot{\mathcal{H}} - \mathcal{H}^2)
- a\alpha_B' - \frac{3}{2}\Omega_m \,,
\end{equation}
with $\alpha \equiv \alpha_K + 6\alpha_B^2$. The vanishing of $\mu_{\Phi,3}$ results from imposing a luminal gravitational-wave speed~\cite{Cusin_2018_de_mg_lss,Cusin_2018_nonlinear_eftofde}.

\section{EFTofLSS and the One-Loop Power Spectrum}
\label{sec:one_loop_spectrum}
To construct the one-loop matter power spectrum, \cref{eq:sec2.effective_euler,eq:sec2.effective_poisson,eq:sec2.effective_continuity} must be solved for $\delta$ up to third order, including counterterms and stochastic contributions. We first discuss the continuity and Euler equations. Afterwards, we present our method to obtain the perturbative solutions, and compare it to the standard Green’s function formalism for solving the density contrast and velocity divergence. Finally, we show how SPT is extended to the EFTofLSS by including counterterms and stochastic contributions from galaxy density fluctuations.

\subsection{Fourier-Space Evolution Equations}
\label{subsec:fourier_space}
The first-order (linear) density contrast obeys $\delta^{(1)}(\vec{k},a) = D(a)\,\delta^{(1)}(\vec{k},a_i)$ for an initial scale factor $a_i$ taken deep in the matter-dominated era, which defines the linear growth factor $D(a)$. Starting from the linear order, one can combine the continuity, Euler and Poisson equations into a second order differential equation for $D(a)$ on sub-horizon scales, satisfying
\begin{equation}
\frac{d^2 D(a)}{d(\ln a)^2} + \left(1+ \frac{\dot{\mathcal{H}}}{\mathcal{H}^2}\right)\frac{dD(a)}{d\ln a} - \frac{3}{2}\mu_\Phi \Omega_m D(a) = 0\,.
\end{equation}
The linear growth factor is scale-independent because $\mu_\Phi=\mu_\Phi(a)$ carries no $k$-dependence in \cref{eq:mu-phi-eqs}; this is a peculiar feature of Vainshtein-screened models and underlies the factorized ansatz for the higher-order kernels introduced below.

The equation has two solutions, $D_{\pm}$, with initial conditions $D_{+}\propto a$ and $D_{-} \propto a^{-3/2}$ in the matter dominated epoch. In the following, we denote the growing-mode solution by $D$. Once modifications to gravity become relevant at late times, $\mu_\Phi$ deviates from unity and the simple power-law behaviour above no longer holds, so the growing- and decaying-mode character of the two solutions is not, in general, preserved beyond matter domination. Throughout this work, we define $D$ as the solution continuously connected to the standard growing mode during matter domination: imposing the standard initial conditions, $D\propto a$ and $dD/d\ln a \propto a$ deep in matter domination, the decaying-mode contribution is negligible by construction and the solution tracks $D_+$ at late times, unless one deliberately chooses a non-standard (fine-tuned) initial velocity that instead enhances the decaying mode. The corresponding linear growth rate is defined as:
\begin{equation}
f \equiv \frac{d\ln D}{d\ln a},
\end{equation}
We now go back to the nonlinear equations. We introduce the rescaled velocity divergence
\begin{equation}
\label{eq:sec3.rescaled_velocity_divergence}
\theta \equiv -\frac{\partial_iv^i}{f\mathcal{H}}.
\end{equation}
and transform to Fourier space. The continuity and Euler equations, respectively \cref{eq:sec2.effective_continuity,eq:sec2.effective_euler}, then take the form~\cite{Cusin:2017wjg}
\begin{align}
    \label{eq:sec2.continuity_fourier}
    a \delta'(\vec{k}, a) - f\theta(\vec{k}, a) &= \int_{\vec{k}_1} \int_{\vec{k}_2} (2\pi)^3 \delta_D(\vec{k} - \vec{k}_{12}) \notag \\
    &\quad \times f\alpha(\vec{k}_1, \vec{k}_2) \theta(\vec{k}_1, a) \delta(\vec{k}_2, a) \\
    \label{eq:sec2.euler_fourier}
    a (f\theta)'(\vec{k}, a) + \left(1 + \frac{ \dot{\mathcal{H}}}{\mathcal{H}^2} \right) f\theta(\vec{k}, a) + \frac{k^2}
    {\mathcal{H}^2} \Phi(\vec{k}, a)
    &= \int_{\vec{k}_1} \int_{\vec{k}_2} (2\pi)^3 \delta_D(\vec{k} - \vec{k}_{12}) \notag \\
    &\quad \times \beta(\vec{k}_1, \vec{k}_2) f^2 \theta(\vec{k}_1, a) \theta(\vec{k}_2, a) \\
    &\quad + \mathcal{H}^{-2} \int d^3x \, e^{i \vec{k} \cdot \vec{x}} \, \partial_i \left( \rho_m^{-1} \partial_j \tau^{ij}_s(\vec{x}, a) \right) \notag,
\end{align}
where $\int_{\vec{k}}\equiv \int\frac{d^3k}{(2\pi)^3}$, $\vec{k}_{1 \dots n} \equiv \sum^n_{i=1} \vec{k}_i$, and the standard dark matter interaction vertices are
\begin{equation}
    \alpha(\vec{q}_1,\vec{q}_2) = 1 + \frac{\vec{q}_1\cdot\vec{q}_2}{q_1^2}\quad\text{and}\quad\beta(\vec{q}_1,\vec{q}_2) = \frac{q_{12}^2 (\vec{q}_1\cdot\vec{q}_2)}{2q_1^2q_2^2}.
\end{equation}
To close the system of equations, the gravitational potential must be related to the matter overdensity through the Poisson equation. In Fourier space, this relation reads~\cite{Cusin:2017mzw,Cusin:2017wjg}
\begin{align}
    \label{eq:sec3.poisson_fourier}
    -\frac{k^2}{\mathcal{H}^2} \Phi(\vec{k}, a) &= \mu_{\Phi} \frac{3 \Omega_m}{2} \delta(\vec{k}, a) \notag \\
    &+ \mu_{\Phi,2} \left( \frac{3 \Omega_m}{2} \right)^2 \int_{\vec{k}_1} \int_{\vec{k}_2} (2\pi)^3 \delta_D(\vec{k} - \vec{k}_{12}) \, \gamma_2(\vec{k}_1, \vec{k}_2) \delta(\vec{k}_1, a) \delta(\vec{k}_2, a) \notag \\
    &+ \mu_{\Phi,3} \left( \frac{3 \Omega_m}{2} \right)^3 \int_{\vec{k}_1} \int_{\vec{k}_2} \int_{\vec{k}_3} (2\pi)^3 \delta_D(\vec{k} - \vec{k}_{123}) \, \gamma_3(\vec{k}_1, \vec{k}_2, \vec{k}_3) \notag \\
    &\qquad \qquad \hspace{3cm} \times \delta(\vec{k}_1, a) \delta(\vec{k}_2, a) \delta(\vec{k}_3, a) \notag \\
    &+ \mu_{\Phi,22} \left( \frac{3 \Omega_m}{2} \right)^3 \int_{\vec{k}_1} \int_{\vec{k}_2} \int_{\vec{q}_1} \int_{\vec{q}_2} (2\pi)^3 \delta_D(\vec{k} - \vec{k}_{12})(2\pi)^3 \delta_D(\vec{k}_2 - \vec{q}_{12}) \notag \\
    &\qquad \qquad \hspace{3cm} \times \gamma_2(\vec{k}_1, \vec{k}_2) \gamma_2(\vec{q}_1, \vec{q}_2) \delta(\vec{k}_1, a) \delta(\vec{q}_1, a) \delta(\vec{q}_2, a) \, .
\end{align}

with the momentum-dependent interaction vertices
\begin{align}
    \gamma_2(\vec{k}_1, \vec{k}_2) &= 1 - \frac{(\vec{k}_1 \cdot \vec{k}_2)^2}{k_1^2 k_2^2} \\
    \gamma_3(\vec{k}_1, \vec{k}_2, \vec{k}_3) &= \frac{1}{k_1^2 k_2^2 k_3^2} \Big(
        k_1^2 k_2^2 k_3^2 + 2(\vec{k}_1 \cdot \vec{k}_2)(\vec{k}_1 \cdot \vec{k}_3)(\vec{k}_2 \cdot \vec{k}_3) \notag \\
        &\qquad - (\vec{k}_1 \cdot \vec{k}_3)^2 k_2^2 - (\vec{k}_2 \cdot \vec{k}_3)^2 k_1^2 - (\vec{k}_1 \cdot \vec{k}_2)^2 k_3^2
    \Big).
\end{align}

\subsection{Perturbative solutions: Green's functions versus ODE approach}

To solve \cref{eq:sec2.continuity_fourier,eq:sec2.euler_fourier} order by order, neglecting momentarily the effective stress tensor on the right-hand side of \cref{eq:sec2.effective_euler}, we write the density contrast and rescaled velocity divergence as a perturbative series~\cite{Cusin:2017wjg}:
\begin{equation}\label{eq:sec2.pertansatz}
\delta(\vec{k},a) = \sum^\infty_{n=1}\delta^{(n)}(\vec{k},a), \qquad  \theta(\vec{k},a) = \sum^\infty_{n=1}\theta^{(n)}(\vec{k},a)\,,
\end{equation}
where $\delta^{(n)}$, $\theta^{(n)}$ are the $n$-th order solutions to the continuity and Euler equations in the absence of the effective stress tensor. The effect of the latter is reinstated in \cref{subsec:galaxy_power_spectrum} as a counterterm contribution to the one-loop galaxy power spectrum.

At each perturbative order $n$, the right-hand sides of \cref{eq:sec2.euler_fourier,eq:sec2.continuity_fourier} act as source terms built from products of lower-order solutions. Since the linear solution satisfies $\delta^{(1)}(\vec{k},a)\propto D(a)$, these source terms carry an overall time dependence $\propto D^n(a)$, multiplied by purely momentum-dependent geometric factors. This structure motivates a factorized ansatz for the $n$-th order solutions:
\begin{align}
\delta^{(n)}(\vec{k},a) = \left(\prod^{n}_{i=1} \int_{\vec{q}_i} \right)(2\pi)^{3}\delta_{D}(\vec{k}-\vec{q}_{1 \dots n})&K_\delta^{(n)}(\vec{q}_1,\dots,\vec{q}_n,a) \delta^{(1)}(\vec{q}_1,a)\dots\delta^{(1)}(\vec{q}_n,a), \nonumber \\
\theta^{(n)}(\vec{k},a) =\left(\prod^{n}_{i=1} \int_{\vec{q}_i} \right)(2\pi)^{3}\delta_{D}(\vec{k}-\vec{q}_{1 \dots n})&K_\theta^{(n)}(\vec{q}_1,\dots,\vec{q}_n,a) \delta^{(1)}(\vec{q}_1,a)\dots\delta^{(1)}(\vec{q}_n,a),
\end{align}
where the kernels $K^{(n)}_\lambda$ $(\lambda \in\left\{\delta,\theta\right\})$ decompose into purely momentum-dependent geometric pieces, built from the interaction vertices $\alpha,\ \beta,\ \gamma_2,\ \gamma_3$, multiplied by purely time-dependent scalar coefficients. These scalar coefficients define the set of time-dependent functions \cite{Donath:2020abv}
\begin{equation}
\label{eq:curly_function_set}
\left\{
\mathcal{G}^\lambda_\sigma,\,
\mathcal{U}^\lambda_\sigma,\,
\mathcal{V}^\lambda_{\sigma \tilde{\sigma}}
\right\},
\end{equation}
with $\sigma,\tilde{\sigma}\in\{1,2\}$ labeling the continuity- and Euler-sourced solutions, respectively. The explicit expressions of these functions are given in \cref{sec:app1}. Each of these functions satisfies a scalar ODE with a smooth time-dependent source, constructed iteratively from the solutions of the previous perturbative order. Substituting the factorized ansatz into \cref{eq:sec2.continuity_fourier,eq:sec2.euler_fourier} and stripping off the momentum-dependent factors yields, at each order, a linear ODE system for the time-dependent coefficients. We solve the $\sigma=1$ and $\sigma=2$ systems separately and linearly combine the results, integrating forward in $a$ from deep in the matter-dominated era.

This direct ODE approach is equivalent to the standard Green's function method \cite{Lewandowski:2016yce}, in which the perturbative solutions are expressed as convolutions of Green's functions $G_\sigma^\lambda(a,\tilde{a})$ with the order-by-order source terms. These Green's functions satisfy
\begin{align}
\label{eq:sec3.Green_ODE}
a\frac{dG^{\delta}_\sigma(a,\tilde{a})}{da} - fG^\theta_\sigma(a,\tilde{a}) &= \lambda_\sigma \delta_D(a-\tilde{a}), \nonumber \\
a\frac{dG^{\theta}_\sigma(a,\tilde{a})}{da}- fG^\theta_\sigma(a,\tilde{a}) + \frac{3\mu_\Phi \Omega_m}{2f}(G^\theta_\sigma(a,\tilde{a}) - G^\delta_\sigma(a,\tilde{a})) &= (1-\lambda_\sigma)\delta_D(a-\tilde{a}),
\end{align}
with $\lambda_1 = 1$ and $\lambda_2 = 0$, and causal boundary conditions
\begin{equation}
G^\delta_\sigma(a,\tilde{a}) =
\begin{dcases}
0, & \tilde{a} > a, \\[4pt]
\dfrac{\lambda_\sigma}{\tilde{a}}, & \tilde{a} = a,
\end{dcases}
\qquad\qquad
G^\theta_\sigma(a,\tilde{a}) =
\begin{dcases}
0, & \tilde{a} > a, \\[4pt]
\dfrac{1-\lambda_\sigma}{\tilde{a}}, & \tilde{a} = a.
\end{dcases}
\end{equation}

By linearity, the solution of \cref{eq:sec3.Green_ODE} convolved with an arbitrary smooth source $S(\tilde{a})$ is equivalent to solving the same ODE with the Dirac delta $\delta_D(a-\tilde{a})$ replaced by $S(a)$, which is what our approach does. Our implementation offers two computational advantages over the convolution approach. First, the source terms require only $D(a)$ from a single prior integration, with no need for the explicit $D_{\pm}$ decomposition required by the Green's function formalism. In contrast, extending the standard convolution method to modified-gravity scenarios requires an explicit identification of the growing mode, making such extensions more difficult (see, e.g.,~\cite{Taule:2024bot}). Second, a single forward ODE integration yields the functions in \cref{eq:curly_function_set} at all scale factors simultaneously, whereas the convolution integral must be recomputed for each target redshift.

\subsection{EFTofLSS one-loop galaxy power spectrum}
\label{subsec:galaxy_power_spectrum}

Thus far, we have neglected the effective stress tensor in \cref{eq:sec2.effective_continuity,eq:sec2.effective_euler}, focusing on the pure perturbation-theory solutions of the matter density and velocity fields. However, observations probe the large-scale structure through galaxies, whose positions are inferred in redshift space and which thus act as biased tracers of the underlying matter distribution. We therefore construct the one-loop galaxy power spectrum using the bias expansion of the tracer density field. Following~\cite{Scoccimarro:1999ed,Senatore:2014vja,Piga:2022mge}, we define the density contrast associated with galaxy density fluctuations as
\begin{equation}
\delta^{(n)}_{g,s}(\vec{k},a) =\left(\prod^{n}_{i=1} \int_{\vec{q}_i} \right)(2\pi)^{3}\delta_{D}(\vec{k}-\vec{q}_{1 \dots n}) Z_n(\vec{q}_1,\dots,\vec{q}_n,a) \delta^{(1)}(\vec{q}_1,a)\dots\delta^{(1)}_{g,s}(\vec{q}_n,a).
\end{equation}
This approach is based on the bias expansion, which was shown using the bootstrap method~\cite{DAmico:2021rdb} to be valid for a wide class of modified gravity models. The functions $Z_1,\, Z_2,\, Z_3$ can be expressed in terms of seven perturbative bias coefficients~\cite{DAmico:2021rdb}. However, as demonstrated in~\cite{DAmico:2021rdb,Piga:2022mge}, for the calculation of the one-loop power spectrum these seven coefficients reduce effectively to four independent parameters, $b_1, \, b_2, \, b_3$ and $b_4$ (see \cref{sec:app2}). The galaxy power spectrum in redshift space, $P_{g,s}(\vec{k},a)$, is defined as
\begin{equation}
\langle \delta_{g,s}(\vec{k},a)\,\delta_{g,s}(\vec{k}^\prime,a) \rangle
= (2\pi)^3 \delta_D(\vec{k}+\vec{k}^\prime)\,P_{g,s}(\vec{k},a).
\end{equation}
It can be used to construct the one-loop galaxy power spectrum in perturbation theory~\cite{Piga:2022mge}:
\begin{equation}
P^{\mathrm{1-loop},\mathrm{PT}}_{g,s}(\vec{k},a) = P_{11}(\vec{k},a) + P_{22}(\vec{k},a)+P_{13}(\vec{k},a),
\end{equation}
with
\begin{align}
P_{11}(\vec{k},a) &= Z_1(\vec{k})^2 P_L(k,a) \\
P_{22}(\vec{k},a) &= 2 \int \frac{d^3 \vec{q}}{(2\pi)^3}\left[Z_2(\vec{k} - \vec{q},\vec{q}) \right]^2 P_L(\vec{q},a)P_L(|\vec{k}-\vec{q}|,a) \\
P_{13}(\vec{k},a) &= 6Z_1(\vec{k})P_L(k,a)\int \frac{d^3 \vec{q}}{(2\pi)^3} Z_3(\vec{k},\vec{q},-\vec{q})P_L(q,a),
\end{align}
where $P_L$ denotes the linear matter power spectrum, defined through
\begin{equation}
\langle \delta^{(1)}(\vec{k},a) \delta^{(1)}(\vec{k}^\prime,a) \rangle = (2\pi)^3 \delta_D(\vec{k}+\vec{k}^\prime)P_L(k,a).
\end{equation}
The result of \cref{eq:sec2.pertansatz} must be corrected for the effects of the UV physics on the long modes, namely the counterterms and stochastic terms resulting from the effective stress tensor of \cref{eq:sec2.effective_euler} and from the bias and redshift-space expansions~\cite{Carrasco:2013mua,Senatore:2014vja,Perko:2016puo,Baldauf:2015xfa}. The full one-loop galaxy power spectrum in redshift space is given by
\begin{equation}
P^{\mathrm{1-loop}}_{g,s}(\vec{k},a) = P^{\mathrm{1-loop},\mathrm{PT}}_{g,s}(\vec{k},a) + P^{\mathrm{CT}}_{g,s}(\vec{k},a) + P^{\epsilon}_{g,s}(\vec{k},a),
\end{equation}
where the counterterm and stochastic contributions read
\begin{align}
P^{\mathrm{CT}}_{g,s}(\vec{k},a) &= 2P_L(k,a)Z_1(\vec{k},a)\left(\frac{k}{k_M}\right)\left(c_{\mathrm{ct}} + c_{r,1}\mu_k^2 + c_{r,2}\mu_k^4 \right) \\
P^{\epsilon}_{g,s}(\vec{k},a) &= \frac{1}{\bar{n}_g}\left(c_{\epsilon,0}+\left(\frac{k}{k_M}\right)^2 \left(c_{\epsilon,1}+f\mu_k^2 c_{\epsilon,2} \right)\right),
\end{align}
where $\bar{n}_g$ denotes the mean galaxy number density, $k_M$ is the characteristic comoving halo scale, and $\mu_k = \hat{k} \cdot \hat{z}$ is the cosine of the angle between the Fourier mode and the line of sight.

\section{Methods}
\label{sec:numerical_implementation}

Having established the theoretical framework for the one-loop galaxy power spectrum in \cref{sec:one_loop_spectrum}, we now describe its implementation and the specifics of our analysis. In \cref{subsec:code,subsec:validation} we present how the perturbation theory equations of \cref{sec:one_loop_spectrum} are incorporated into \pb and validate the resulting time-dependent functions against existing methods. In \cref{subsec:nbody} we present a comparison to $N$-body simulations of \texttt{PySCo-EFT} and \texttt{ECOSMOG-EFT} for a $\alpha_i\propto\Omega_{\rm DE}(a)$ model. We then specify the class of modified gravity models considered in this work (\cref{subsec:model}), together with the parameter space, priors, and observational data used in our inference (\cref{subsec:parameters_and_priors}).

\subsection{The code}
\label{subsec:code}
We use the code \pb \footnote{Public GitHub repository available at \href{https://github.com/pierrexyz/PyBird}{https://github.com/pierrexyz/PyBird}} \cite{DAmico:2020kxu}, written in Python, to compute the multipoles of the power spectrum of biased tracers in redshift space. The code is based on Refs. \cite{DAmico:2019fhj,Perko:2016puo} and implements a theoretical framework built on EFTofLSS together with a perturbative bias expansion, as presented in \cite{Donath:2020abv}.

In the previous sections we derived the perturbation theory equations relevant for scale-independent theories of modified gravity. We incorporate these into \pb by modifying its computation of the Green’s functions with our direct ODE approach. Since the modifications to the perturbative dynamics enter through the background evolution (\eg via $D(a)$ and $\mathcal{H}(a)$) and the interaction vertices (through $\mu_{\Phi,\{2,22\}}$), we introduce a new API that allows these quantities to be supplied as user-defined functions. We further implement a procedure to construct the vertices directly from the $\alpha$-parameters following \cref{eq:mu-phi-eqs}. This design makes our implementation flexible and readily applicable to a broad class of scale-independent modified-gravity models. In particular, our approach can interface with Einstein–Boltzmann solvers, such as $\mathcal{H}$–\texttt{EFTCAMB}, by using spline interpolations constructed from their grid outputs as inputs to our code.

\subsection{Numerical consistency tests and wall time comparison}
\label{subsec:validation}

\begin{figure}[H]
    \centering
    \includegraphics[width=0.8\linewidth]{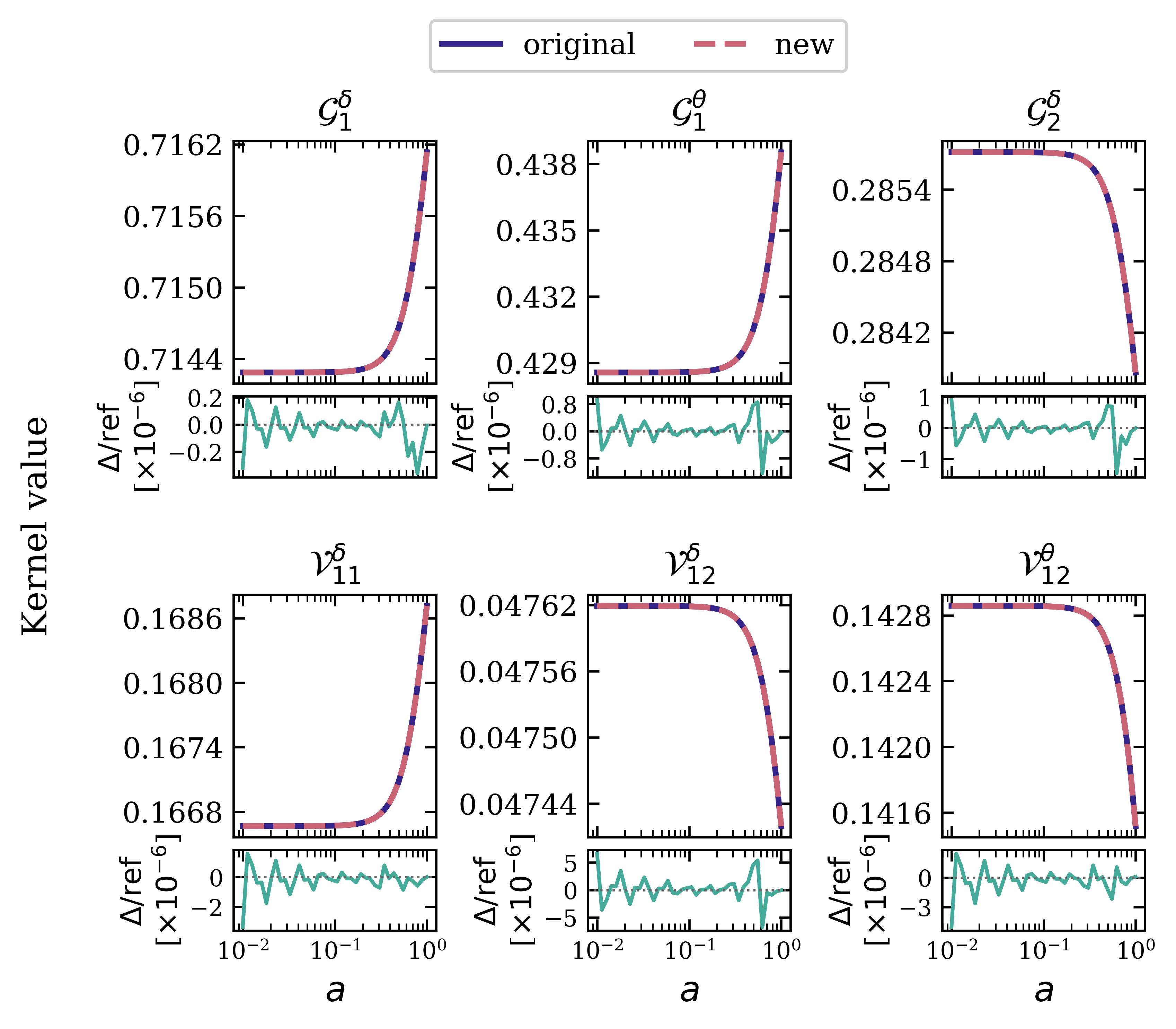}
    \caption{Comparison of the time-dependent functions computed in a flat $\Lambda$CDM model with $\Omega_{m,0}=0.3$, using the original Green's function method implemented in \texttt{\pb} (indigo, solid) and our direct ODE integration approach (rose, dashed). The fractional difference (lower panels) remains at the level of $\mathcal{O}(10^{-6})$ over the full range of scale factors.}
    \label{fig:compare_implementation_LCDM_minimal}
\end{figure}

In this subsection, we perform numerical comparisons against \texttt{PyBird} and an independent custom implementation to verify the numerical consistency of our pipeline. Note that for the computation of the one-loop power spectrum, only six functions in \cref{eq:curly_function_set} are required, namely $\mathcal{G}_1^{\{\delta,\theta\}},\ \mathcal{G}_2^\delta,\ \mathcal{V}_{11}^\delta,\ \mathcal{V}_{12}^{\{\delta,\theta\}}$. We compare these functions obtained with our novel ODE method to those computed using the original
\texttt{PyBird} Green's function approach for the models available in the
standard \texttt{PyBird} implementation, namely $\Lambda$CDM, $w$CDM, and
Quintessence~\cite{D_Amico_2024}. In all these cases, we have
$\mu_{\Phi}=1$ and $\mu_{\Phi,2}=\mu_{\Phi,22}=0$. The latter two models allow us to test our implementation with a modified background evolution while keeping the Poisson equation fixed to its $\Lambda$CDM form. For all three cases, we find excellent agreement, with a fractional difference of $\mathcal{O}\left(10^{-6}\right)$ (see \eg ~\cref{fig:compare_implementation_LCDM_minimal}).

The original \texttt{\pb} code does not include modified gravity theories with $\mu_{\Phi}\neq1$ and/or $\mu_{\Phi,2},\, \mu_{\Phi,22}\neq0$. To test our method in such modified gravity scenarios, we performed a consistency check for two representative models: nDGP~\cite{Dvali:2000hr,Koyama:2007ih} \footnote{For nDGP, see e.g. Ref.~\cite{Piga:2022mge} for the equations of $\mu_\Phi$ and $\mu_{\Phi,\{2,22,3\}}$.}, with a $\Lambda$CDM background and modified perturbations, and Cubic Galileon (CG) with the tracker solution~\cite{DeFelice:2010pv}, with modified background evolution and perturbations. We independently implemented the standard Green's function method and compared its time-dependent functions with those obtained using the novel direct ODE method. The two approaches agree to high precision, with fractional differences at the $\mathcal{O}(10^{-6})$ level, as expected from the mathematical equivalence of the two methods.

Finally, for all tested models we find a similar level of agreement between the two approaches for the remaining functions of the set defined in \cref{eq:curly_function_set}, beyond the minimal six required for the one-loop power spectrum. While these additional functions are not needed for the one-loop calculations presented here, they are required for higher-order observables, such as the bispectrum and the two-loop matter power spectrum, providing an additional consistency check for future applications.

\begin{figure}[H]
    \centering
    \includegraphics[width=0.8\linewidth]{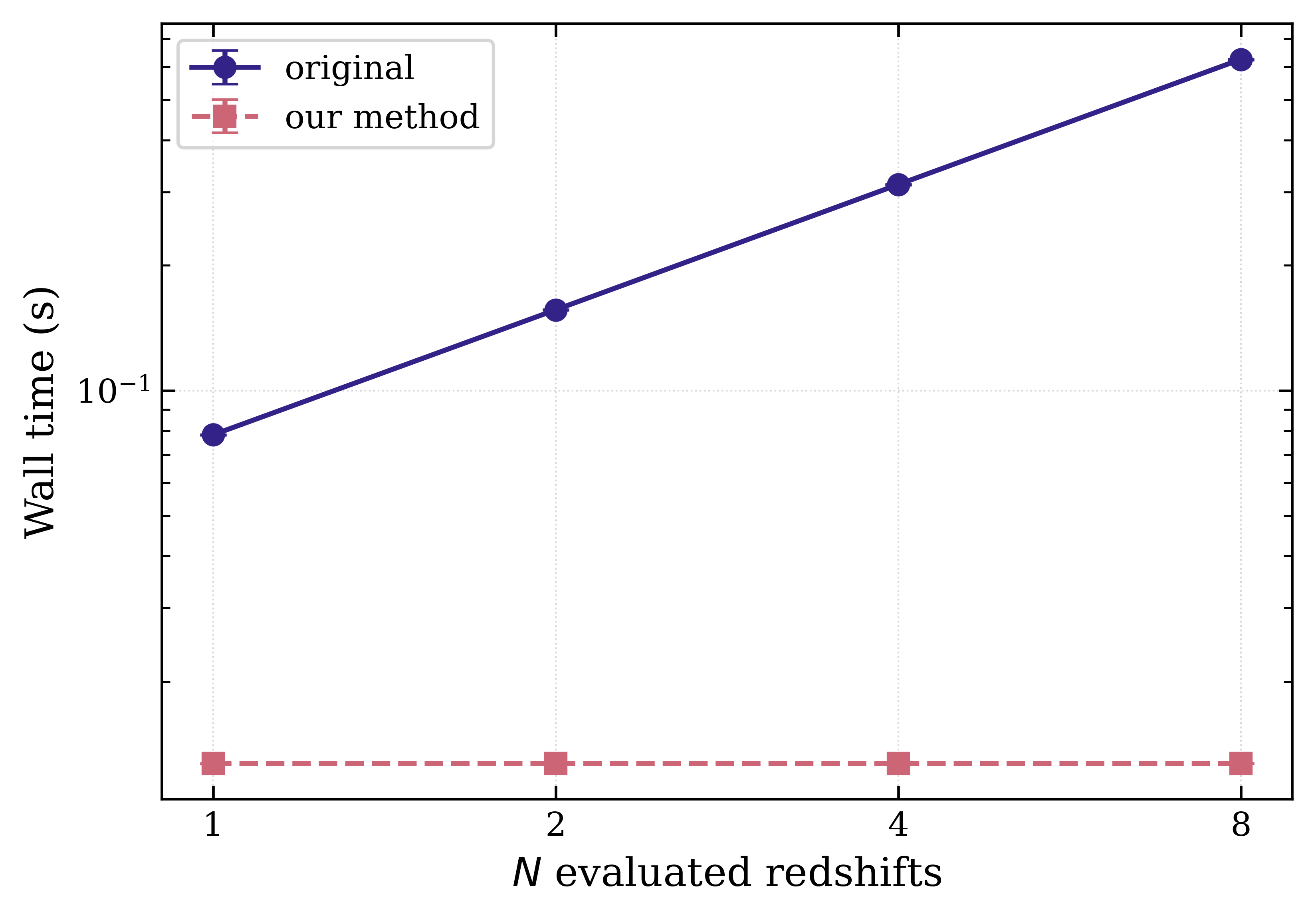}
    \caption{Average wall time (in seconds) for computing the minimal set $\left\{\mathcal{G}_1^{\{\delta,\theta\}},\ \mathcal{G}_2^\delta,\ \mathcal{V}_{11}^\delta,\ \mathcal{V}_{12}^{\{\delta,\theta\}}\right\}$ as a function of the number of evaluated redshifts in a flat $\Lambda$CDM model with $\Omega_{m,0}=0.3$. Standard \texttt{\pb} (indigo, solid) is compared with our direct ODE integration approach (rose, dashed). Error bars indicate the standard deviation over 50 runs.}
    \label{fig:compare_timings}
\end{figure}

Comparing the computation time for different numbers of target redshifts, we find that our method evaluates faster and scales independently of the number of target redshifts, while the Green’s function convolution must be repeated per redshift (\cref{fig:compare_timings}). The relatively small cost of evaluating additional redshifts after initialization is beneficial for likelihood analyses, where each model evaluation may require predictions at multiple redshifts.
In the wall time comparison, we considered only the minimal set of functions required for the one-loop power spectrum. Even when evaluating all functions in \cref{eq:curly_function_set}, the ODE evaluation time remains roughly unchanged at $\sim10^{-2}\ \mathrm{s}$. By contrast, the Green's function approach becomes approximately three times more expensive than reported here, since each additional function requires a separate convolution integral.

\Cref{fig:compare_timings} isolates the cost of the kernel evaluation itself. Considering the full one-loop calculation, including the loop integrals shared by both approaches, computing the monopole and quadrupole at a single redshift takes $\sim 300$--$400\,\mathrm{ms}$ with our method compared to $\sim 400$--$450\,\mathrm{ms}$ with the standard implementation, on a single core without \texttt{JAX}-based acceleration \cite[see][for recent \LCDM-optimized timings]{reeves2026pybirdjaxacceleratedinferencelargescale}. Because the loop-integral computation dominates this total and is common to both approaches, the resulting $\sim 20$--$25\%$ speedup at the level of the full one-loop spectrum is far more modest than the order-of-magnitude difference in \cref{fig:compare_timings}, which reflects the kernel-evaluation cost alone. The advantage of our approach becomes more relevant when evaluating multiple redshifts within a fixed cosmology: since the growth ODE is solved once, our method saves $\sim 100\,\mathrm{ms}$ per additional redshift relative to the standard implementation, which must reconstruct its Green's function at every redshift; this saving accumulates linearly with the number of target redshifts.

\subsection{Comparison to N-body simulations}
\label{subsec:nbody}

The tests above verify the internal numerical consistency of our direct ODE method against the standard Green's function formalism, but do not test the accuracy of the underlying one-loop, Vainshtein-screened perturbation theory against fully nonlinear structure formation. To assess this, we compare our one-loop matter power spectrum against the $N$-body simulation suite of \cite{Ganjoo:2026pysco}, run with the \texttt{PySCo-EFT} and \texttt{ECOSMOG-EFT} codes for the $\alpha_i\propto\Omega_{\rm DE}(a)$ parametrization used throughout this work (with $\alpha_T=0$), for several representative $\{\alpha_B^0,\alpha_M^0\}$ combinations at $z=0$ (see Figs.~9 and~10 of Ref.~\cite{Ganjoo:2026pysco}), where $\alpha_i^0\equiv\alpha_i(a=1)$ denotes the physical present-day value. This provides an independent, nonlinear confirmation of the one-loop Vainshtein-screened matter power spectrum implemented in our pipeline, complementing the internal consistency checks of \cref{subsec:validation}.

We adopt the cosmology of \cite{Ganjoo:2026pysco} ($h=0.6803$, $\Omega_{m,0}=0.3071$, $n_s=0.9660$, $\sigma_8^{\Lambda\mathrm{CDM}}=0.8220$), fixing $\omega_b$ to the Planck 2018 value since it is not separately reported. For each $\{\alpha_B^0,\alpha_M^0\}$ pair we compute the real-space, unbiased one-loop matter power spectrum with our pipeline and fit the single counterterm coefficient $c_{\mathrm{ct}}$ against the $N$-body $P(k)$ for $k<0.3\,h\,\mathrm{Mpc}^{-1}$, following standard practice in one-loop-vs-$N$-body comparisons; all other EFT coefficients are set to zero.

\begin{figure}[H]
    \centering
    \includegraphics[width=0.8\linewidth]{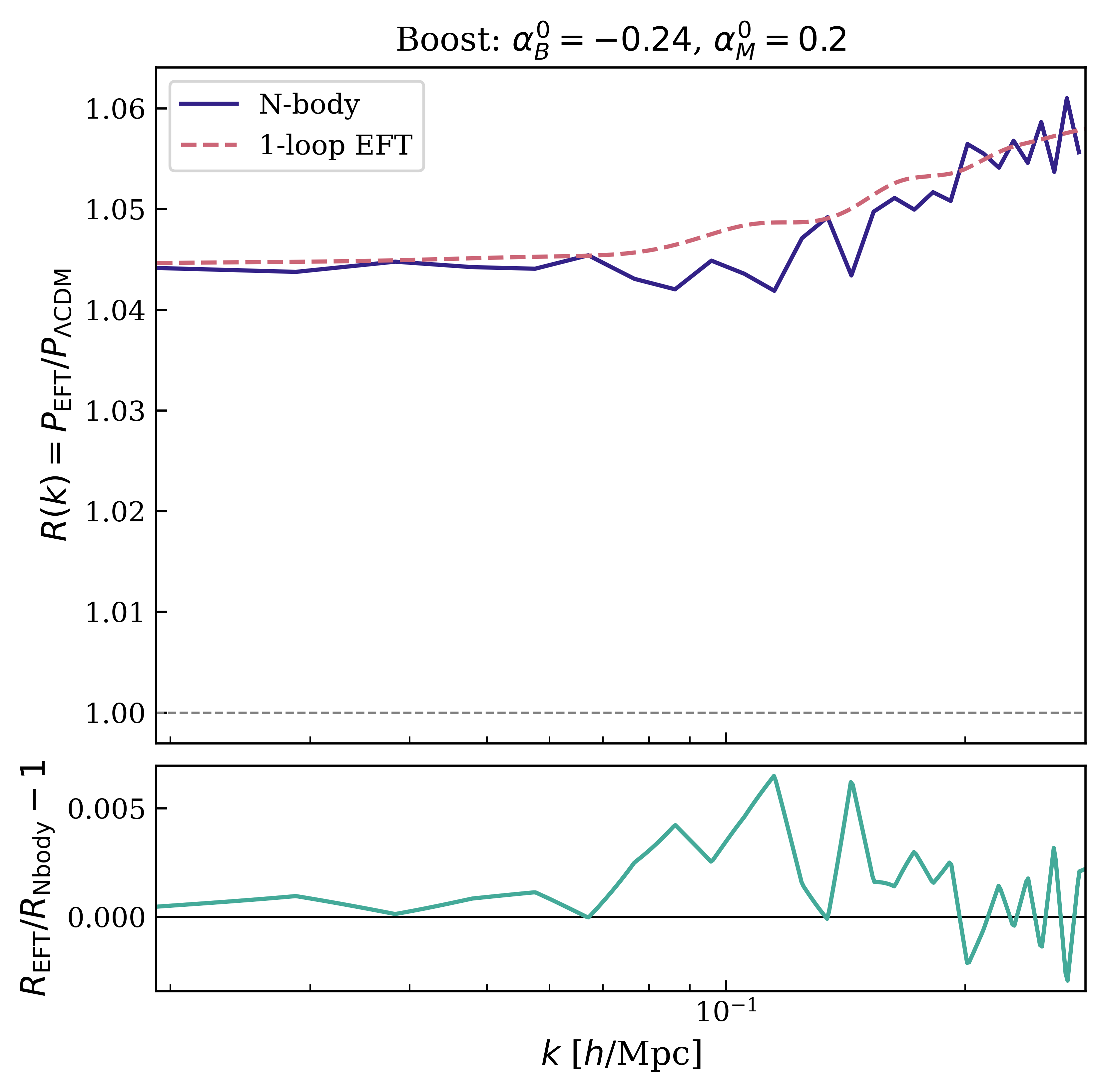}
    \caption{Boost $R(k)\equiv P_{\rm EFT}/P_{\Lambda\mathrm{CDM}}$ at $z=0$ for $\alpha_B^0=-0.24$, $\alpha_M^0=0.2$, comparing the one-loop prediction (rose) to the $N$-body measurement of \cite{Ganjoo:2026pysco} (indigo). Lower panel: fractional residual between the two boosts.}
    \label{fig:nbody_boost}
\end{figure}

Rather than comparing the raw power spectra, which are dominated by the cosmic variance of the modest simulation volume (and for which the single counterterm coefficient $c_{\rm ct}$ is fit directly to the $N$-body $P(k)$), we compare the boost $R(k)\equiv P_{\rm EFT}/P_{\Lambda\mathrm{CDM}}$, for which most of the cosmic variance common to the modified-gravity and $\Lambda$CDM runs cancels. \Cref{fig:nbody_boost} shows this comparison for $\alpha_B^0=-0.24$, $\alpha_M^0=0.2$, values representative of the amplitude range probed by our posterior analysis (\cref{sec:results.constraints_on_amplitudes}): the one-loop prediction agrees with the $N$-body boost at the sub-percent level over $0.1\lesssim k \lesssim 0.3\,h\,\mathrm{Mpc}^{-1}$. We have confirmed that the other EFTofDE parameter combinations give similar agreement. In addition, we compared the EFT and \(\Lambda\)CDM nonlinear matter power spectra individually against the \(N\)-body simulations. In all cases, the predictions are consistent with the simulation results within the estimated cosmic variance from the simulation box size.

\subsection{Model specification}
\label{subsec:model}

The pipeline described in the previous subsection is broadly applicable to the class of luminal Horndeski theories and can interface with any code capable of computing their linear cosmological evolution. To demonstrate its application, we specialize to representative realizations implemented in $\mathcal{H}$--\texttt{EFTCAMB}. Specifically, we perform a complete cosmological analysis, including full-shape large-scale-structure data, sampling the parameter space associated with selected parametrizations of the $\alpha_i$ functions. In addition, we illustrate the resulting one-loop galaxy power spectrum for a selection of covariant models.

We start with a phenomenological parametrization of the time-dependent functions $\alpha_i(a)$ \cite{Bellini:2014fua} (see also \cref{eq:sec2.alphas}), where we remind the reader that we restrict to luminal theories by setting $\alpha_{T}=0$. We  explore two commonly used time dependencies for the remaining free functions $\alpha_B$, $\alpha_M$, and $\alpha_{K}$:
\begin{align}
\alpha_i(a) = c_i \, a^3,
\end{align}
and
\begin{align}
\alpha_i(a) = c_i \, \Omega_{\rm DE}(a),
\end{align}
where $c_i$ are constant coefficients, with $i\in\{B,M,K\}$. In both cases, we fix the background dark energy equation of state to $w = -1$, corresponding to a \(\Lambda\)CDM background expansion history. These parameterizations ensure that deviations from general relativity become relevant only at late times and vanish in the early Universe. Since kineticity has a negligible impact on the large-scale structure observables considered in this work, we fix $c_{K}=0.01$ to a small, nonzero value for numerical stability; the precise choice does not affect our results.

As concrete examples of covariant modified-gravity models, we consider the cubic Galileon with tracker solution~\cite{DeFelice:2010pv}, for which the background and perturbation dynamics are fully fixed once the standard cosmological parameters are specified, and nDGP~\cite{Dvali:2000hr,Koyama:2007ih}, which retains a $\Lambda$CDM background but modifies the perturbation sector through $\mu_\Phi$ and $\mu_{\Phi,\{2,22\}}$. We use them in \cref{sec:results.showcase} to demonstrate the flexibility of our approach across the space of luminal Horndeski theories, computing representative one-loop power spectrum predictions at fixed fiducial parameters rather than performing dedicated parameter inference for each.

The validity of the QSA requires the relevant modes to remain well within the sound horizon~\cite{Sawicki:2015zya}, i.e.\ $k/(aH) \gg c_s^{-1}$, for all times and scales considered in our analysis. To ensure this, we impose the condition $c_s \gtrsim 0.1$~\cite{Sawicki:2015zya,Taule:2024bot}. We also require the absence of ghosts and gradient instabilities to ensure the physical viability of the models considered~\cite{Frusciante:2016xoj,DeFelice:2016ucp}.

\subsection{Parameters, priors, and data}
\label{subsec:parameters_and_priors}

\paragraph{Bias and nuisance parameters.}
Following \cite{DAmico:2019fhj}, we reparametrize the second-order bias coefficients as
\begin{equation}
    c_2\equiv\frac{1}{\sqrt{2}}(b_2+b_4),\qquad c_4\equiv\frac{1}{\sqrt{2}}(b_2-b_4).
\end{equation}
Since $c_4$ enters the signal primarily through the hexadecapole, to which our data have limited sensitivity, we fix $c_4 = 0$. Our free nuisance parameters are therefore
\begin{equation}
    \left\{b_{1},\, c_{2},\, b_{3},\, c_{ \rm ct},\, c_{r,1},\, c_{\epsilon,0},\, c_{\epsilon,1},\, c_{\epsilon,2}\right\},
\end{equation}
where $b_3$ and all counter- and stochastic terms are analytically marginalized over, with priors as in \cite{Piga:2022mge}.

\paragraph{Cosmological and modified-gravity parameters.}
We sample the six standard cosmological parameters
\begin{equation}
    \left\{A_{s},\, n_{s},\, \omega_{b},\, \omega_{c},\, H_{0},\, \tau \right\}.
\end{equation}
For the EFTofDE parametrizations we additionally vary $\{c_B, c_M\}$. Prior distributions for all sampled parameters are listed in \cref{tab:priors}.

\begin{table}[htbp]
    \centering
    \caption{Prior distributions for sampled parameters. All priors are flat (uniform) unless otherwise noted.
    }
    \label{tab:priors}
    \renewcommand{\arraystretch}{1.2}
    \begin{tabular}{lll}
        \hline\hline
        Parameter & Description & Prior \\
        \hline
        $b_{1}$              & Linear bias            & $[-5,\, 5]$ \\
        $c_{2}$              & Second-order bias      & $[-4,\, 4]$ \\
        \hline
        $c_B,\, c_M$         & EFTofDE amplitudes     & $[-5,\, 5]$ \\
        $c_K$                & Kineticity (fixed)     & $0.01$ \\
        \hline
        $\log(10^{10} A_s)$  & Scalar amplitude       & $[1.61,\, 3.91]$ \\
        $n_s$                & Spectral index         & $[0.8,\, 1.2]$ \\
        $H_0$                & Hubble constant        & $[20,\, 100]$ \\
        $\omega_b$           & Baryon density         & $\mathcal{N}(0.0222,\, 0.0005)$ \\
        $\omega_c$           & CDM density            & $[0.001,\, 0.99]$ \\
        $\tau$               & Optical depth          & $[0.04,\, 0.2]$ \\
        \hline\hline
    \end{tabular}
\end{table}

\paragraph{Data.}
We consider three dataset combinations: \textbf{CMB} alone, \textbf{CMB+FS}, and the full joint \textbf{CMB+BAO+FS}. The individual datasets are:
\begin{itemize}
    \item \textbf{CMB}: Planck Commander low-$\ell$ TT \cite{planck2020} and Lollipop low-$\ell$ EE likelihoods, together with the HiLLiPoP \cite{Tristram2023} TTTEEE likelihood at high $\ell$.
    \item \textbf{BAO}: All tracers from DESI DR2 \cite{DESI:2025zgx,DESI:2025zpo}, interfaced via Cobaya \cite{Torrado:2020dgo}.
    \item \textbf{FS}: Post-reconstructed galaxy power spectrum multipoles from BOSS DR12 \cite{Gil_Mar_n_2016}. Specifically, we analyze the monopole and quadrupole of the redshift-space power spectrum for the North and South Galactic Caps, for the redshift bins $z=0.32$ and $z=0.57$. We include modes up to $k_{\mathrm{max}}=0.23\,h/\mathrm{Mpc}$.
\end{itemize}

\section{Results}
\label{sec:results}
In this section, we first illustrate the breadth of modified-gravity models accessible with our pipeline (\cref{sec:results.showcase}), before presenting our constraints on the EFTofDE amplitude parametrizations of \cref{sec:numerical_implementation}, the only models for which we perform full posterior inference in this work. We further discuss projection effects, and carry out a comparison between the adoption of the EdS approximation and our full treatment of the kernels.

\subsection{A showcase across modified-gravity models}
\label{sec:results.showcase}
Before turning to parameter inference, we illustrate the scope of our pipeline across a range of modified-gravity theories. As shown in \cref{subsec:validation}, a single one-loop monopole-and-quadrupole evaluation costs a few hundred milliseconds regardless of the underlying gravity model, so it is computationally inexpensive to survey predictions for a range of models even though a dedicated MCMC analysis for each remains comparatively expensive.

\begin{figure}[H]
    \centering
    \includegraphics[width=0.9\linewidth]{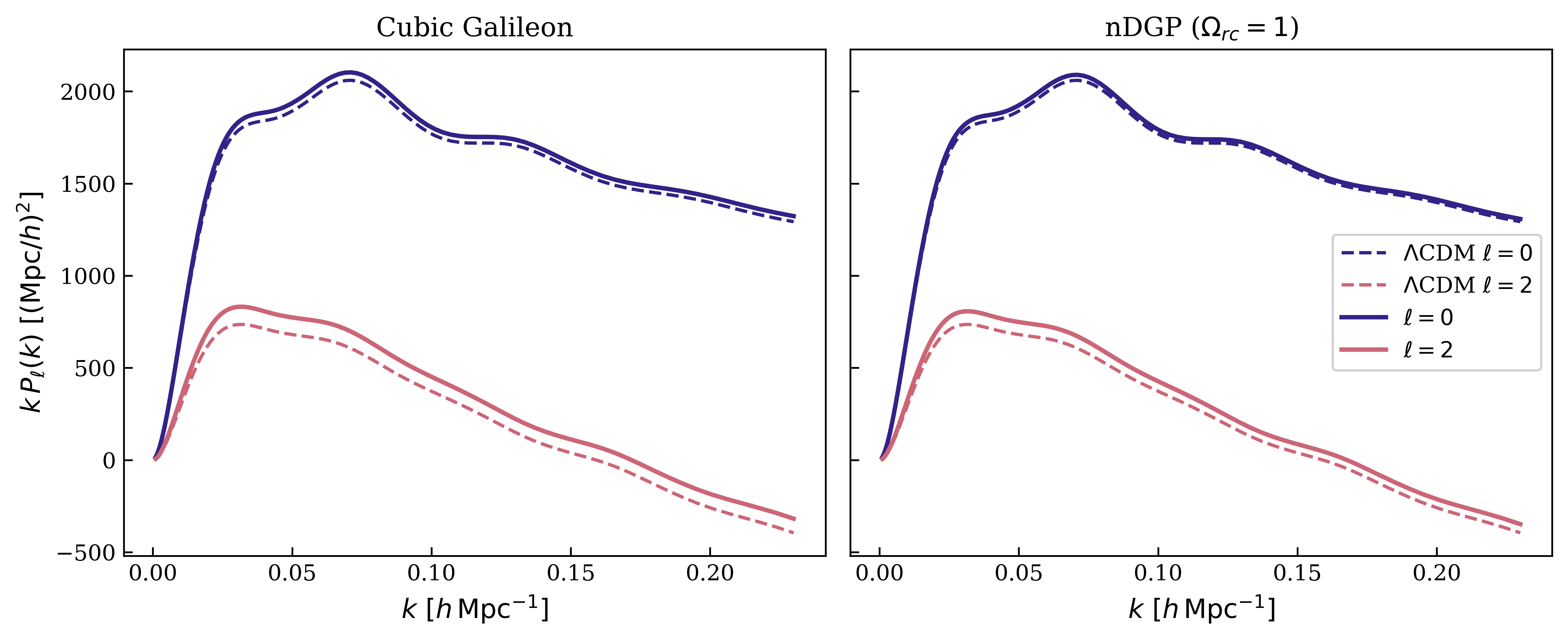}
    \caption{Monopole (indigo) and quadrupole (rose) of the one-loop redshift-space galaxy power spectrum at $z=0.57$ for the cubic Galileon (left) and nDGP with $\Omega_{rc}=1$ (right), each compared to $\Lambda$CDM (dashed) at the same fixed EFT and nuisance parameters.}
    \label{fig:showcase}
\end{figure}

\Cref{fig:showcase} shows the one-loop redshift-space power spectrum multipoles at $z=0.57$ for the two models described in \cref{subsec:model}: the cubic Galileon with tracker solution~\cite{DeFelice:2010pv}, and nDGP~\cite{Dvali:2000hr,Koyama:2007ih}. EFT nuisance parameters are fixed to representative BOSS DR12 best-fit values throughout. These predictions demonstrate that our pipeline extends, at essentially no extra implementation cost, from the phenomenological amplitude parametrizations to any luminal Vainshtein-screened Horndeski model.

\subsection{Constraints on the EFTofDE amplitudes}
\label{sec:results.constraints_on_amplitudes}

\begin{figure}[]
    \centering
    \includegraphics[width=0.47\linewidth]{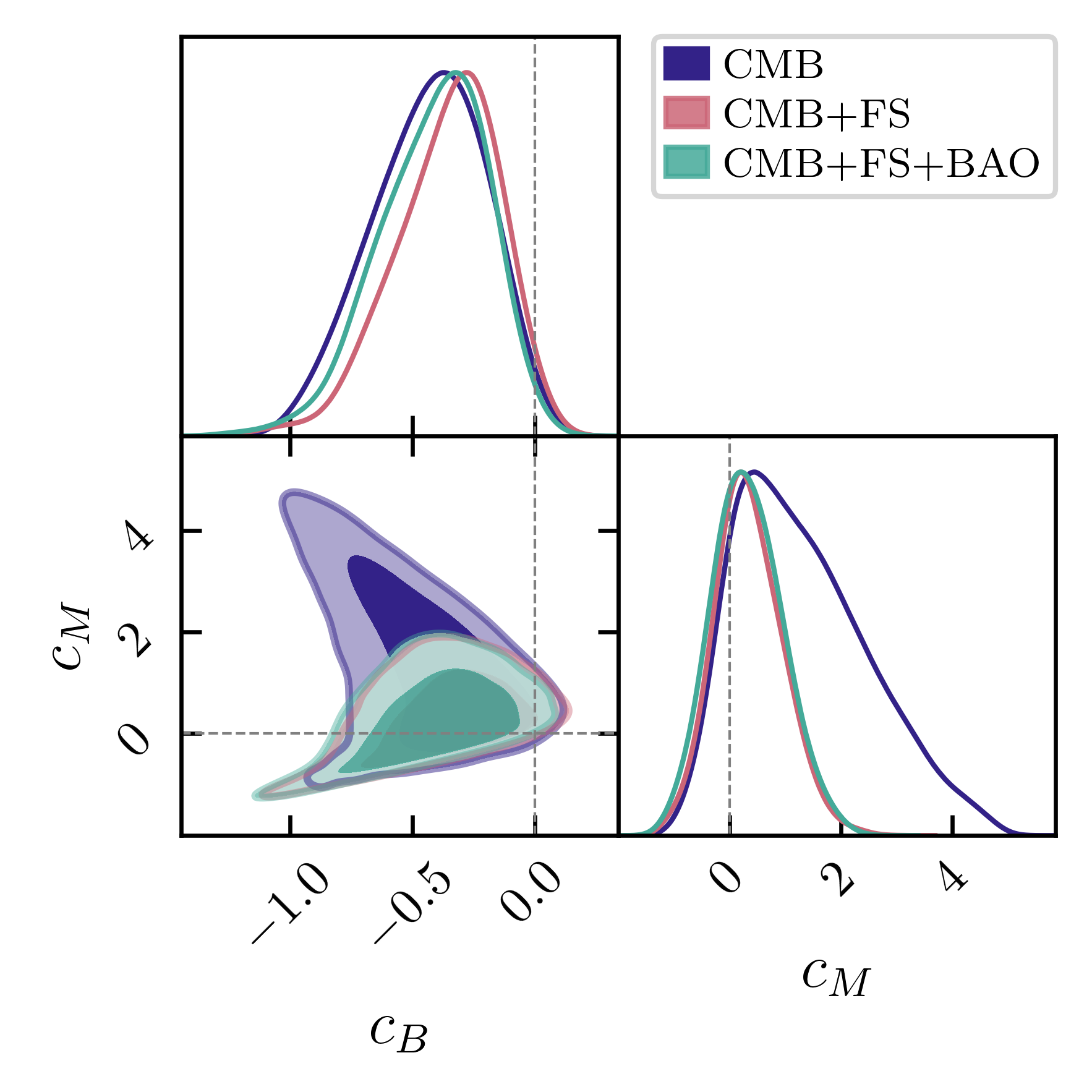}
    \includegraphics[width=0.47\linewidth]{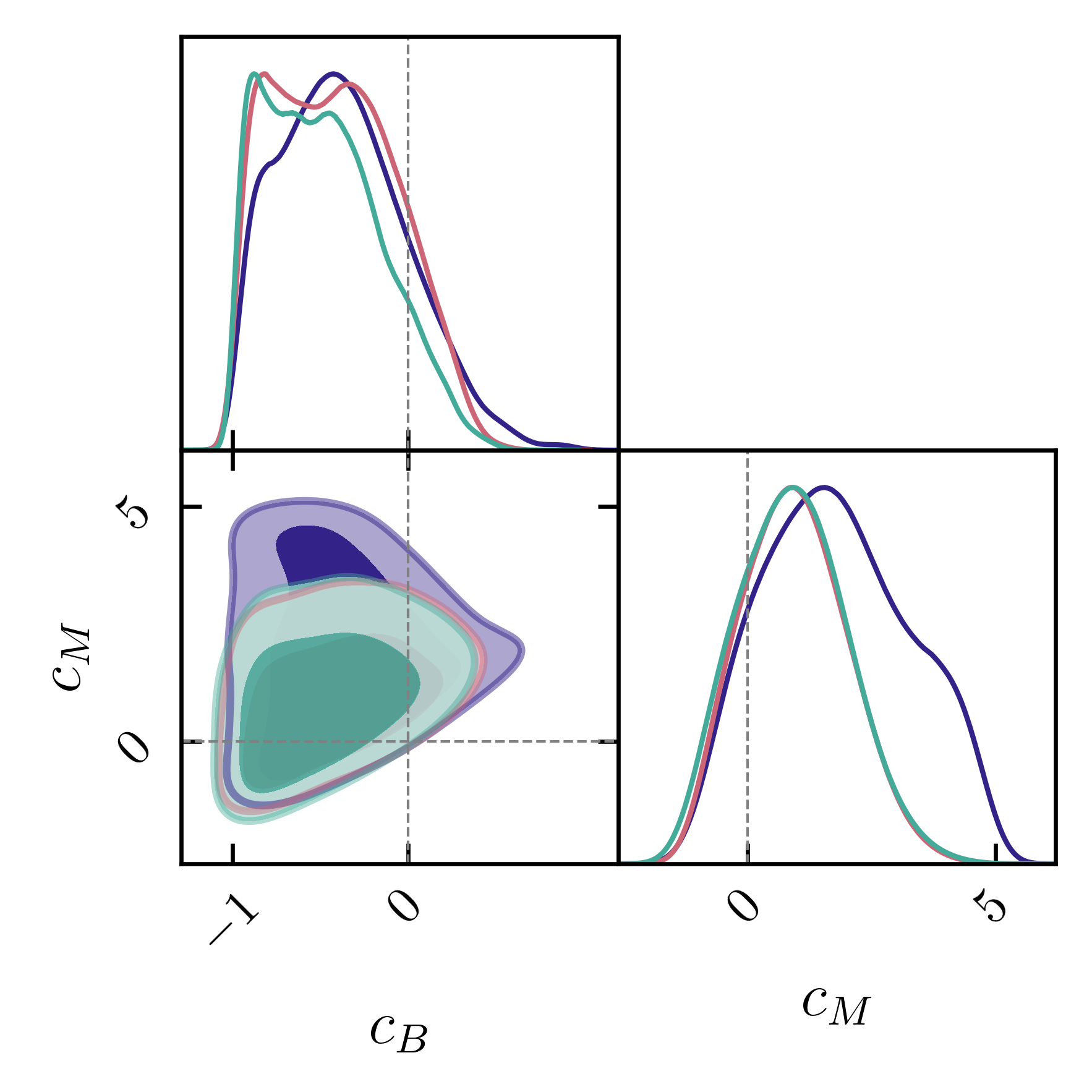}
    \caption{Marginalized 68\% and 95\% CL contours for the EFTofDE amplitude parameters $c_{B}$ and $c_{M}$, for the $\alpha_i=c_i\,\Omega_{\rm DE}$ (left) and $\alpha_i=c_i\,a^{3}$ (right) parametrizations. Indigo, rose, and teal contours show to the CMB-only, CMB+FS, and CMB+FS+BAO constraints, respectively. Dashed lines indicate $c_{B}=c_{M}=0$.}
    \label{fig:alpha_constraints}
\end{figure}

\Cref{fig:alpha_constraints} shows the marginalized 68\% and 95\% CL contours for the modified gravity parameters $c_{B}$ and $c_{M}$ for both parametrizations, comparing CMB-only, CMB+FS, and CMB+BAO+FS constraints.
For the $\alpha_i\propto\Omega_{\rm DE}$ case, CMB data alone indicate a preference for negative braiding, with $c_{B}=-0.43_{-0.21}^{+0.27}$. This preference is consistent with the ISW-mechanism discussed in \cite{Taule:2024bot}, where late-time modifications to $\mu_\Phi$ alter the decay of the Weyl potential sourcing low-$\ell$ TT. By contrast, $c_{M}$ is only weakly constrained by CMB alone, with the posterior extending to large positive values. The lower boundary for $c_{M}$ is driven by the onset of gradient instabilities. Adding the BOSS full-shape galaxy power spectrum and DESI BAO data substantially tightens the constraints on $c_{M}$, yielding $c_{B}=-0.41^{+0.25}_{-0.19}$ and $c_{M}=0.32^{+0.61}_{-0.70}$ at 68\% CL, consistent with GR within the 95\% credible interval.

For the $\alpha_i\propto a^{3}$ parametrization, the qualitative picture is similar, with a preference for $c_{B}<0$ and a wide $c_{M}$ posterior from CMB alone. The CMB+FS+BAO combination again tightens the $c_{M}$ constraint. The results for both parametrisations, overall consistent with GR at the 95\% CL, are denoted in \cref{tab:post_means}.
\begin{table}[]
    \centering
    \begin{tabular}{lcccc}
        \hline\hline
                        & \multicolumn{2}{c}{$\propto \Omega_\mathrm{DE}$} & \multicolumn{2}{c}{$\propto a^3$} \\
        \cline{2-3} \cline{4-5}
                        & $c_B$ & $c_M$ & $c_B$ & $c_M$ \\
        \hline
        CMB             & $-0.431^{+0.270}_{-0.212}$ & $1.314^{+0.837}_{-1.528}$ & $-0.373^{+0.263}_{-0.482}$ & $1.811^{+1.380}_{-1.878}$ \\
        CMB + FS        & $-0.357^{+0.256}_{-0.174}$ & $0.343^{+0.560}_{-0.668}$ & $-0.419^{+0.259}_{-0.487}$ & $0.941^{+1.005}_{-1.184}$ \\
        CMB + FS + BAO  & $-0.414^{+0.260}_{-0.187}$ & $0.318^{+0.608}_{-0.696}$ & $-0.487^{+0.203}_{-0.462}$ & $0.893^{+1.122}_{-1.196}$ \\
        \hline\hline
    \end{tabular}
    \caption{Marginalized posterior means and 68\% credible intervals for $c_B$ and $c_M$, for both the $\propto\Omega_\mathrm{DE}$ and $\propto a^3$ parametrizations, using the CMB, CMB+FS, and CMB+FS+BAO data combinations shown in \cref{fig:alpha_constraints}.}
    \label{tab:post_means}
\end{table}

We do note a sharp cutoff at $c_{B}=-1$ in the posteriors for the $\propto a^{3}$ parametrization. This boundary arises from a singularity in the $00$ component of the Einstein equations at $\alpha_{B}=-1$ \cite[see \eg][Eq.(A.2)]{Ye:2026qqf}. Since for the cubic parametrization $c_{B}\equiv\alpha_{B}(a=1)$, this singularity maps directly onto the $c_{B}=-1$ boundary visible in the contours. For the $\propto\Omega_{\rm DE}$ parametrization, the same singularity occurs only at $c_{B}=-1/\Omega_{\rm DE}(a=1)<-1$, which lies outside the plotted range and therefore does not affect those posteriors. For both parametrizations, the posterior constraints of the cosmological parameters are shown in \cref{app:full_post_constraints}.

\begin{figure}[]
    \centering
    \includegraphics[width=0.9\linewidth]{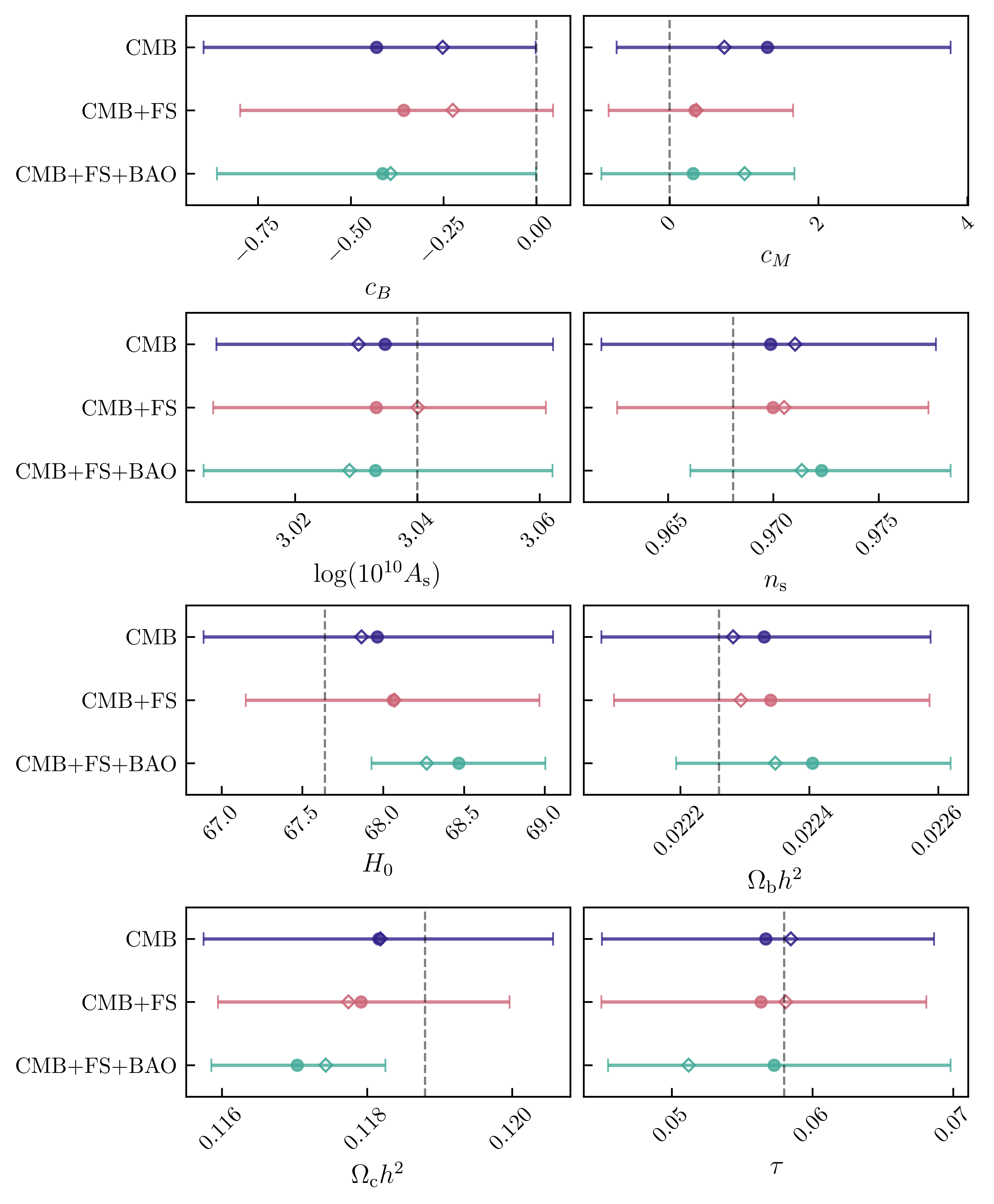}
    \caption{Constraints on the cosmological and EFTofDE parameters for the $\Omega_{\rm DE}$ parametrization. Diamonds show the maximum a posteriori (MAP) values, circles the posterior means, and bars the 95\% confidence intervals. Dashed lines indicate the $\Lambda$CDM fiducial values. Results are shown for the CMB, CMB+FS, and CMB+FS+BAO combinations.}
    \label{fig:ODE_projection}
\end{figure}

\subsection{Projection effects}
In the EFTofLSS framework, projection effects (shifts in the Maximum A Posteriori (MAP) compared to posterior mean for a given parameter) may occur when nuisance parameters, such as bias, stochastic terms, and counter-terms, are partially degenerate with cosmological parameters ~\cite[see \eg][for detailed discussions]{Adame_2025}. In FS-only analyses, these projection effects may be alleviated by including strong CMB-informed priors on $A_s$ and $n_s$ \cite{Moretti_2023}, or by using a different basis for the EFTofLSS nuisance parameters \cite{tsedrik2025simplewaymeasureevolving}. In our analysis, however, FS always appears in combination with CMB probes. The CMB  tightly constrains $A_s$, breaking possible degeneracies, and we therefore expect no strong projection effects. We compare the MAP obtained with \texttt{Py-BOBYQA} ~\cite{Powell2009,cartis2018improvingflexibilityrobustnessmodelbased,Cartis_2021} with our posterior means in \cref{fig:ODE_projection}, finding good agreement between the two, confirming that projection effects are not a significant concern in our analysis.

\subsection{EdS versus exact time evolution}

\begin{figure}[H]
    \centering
    \includegraphics[width=0.8\linewidth]{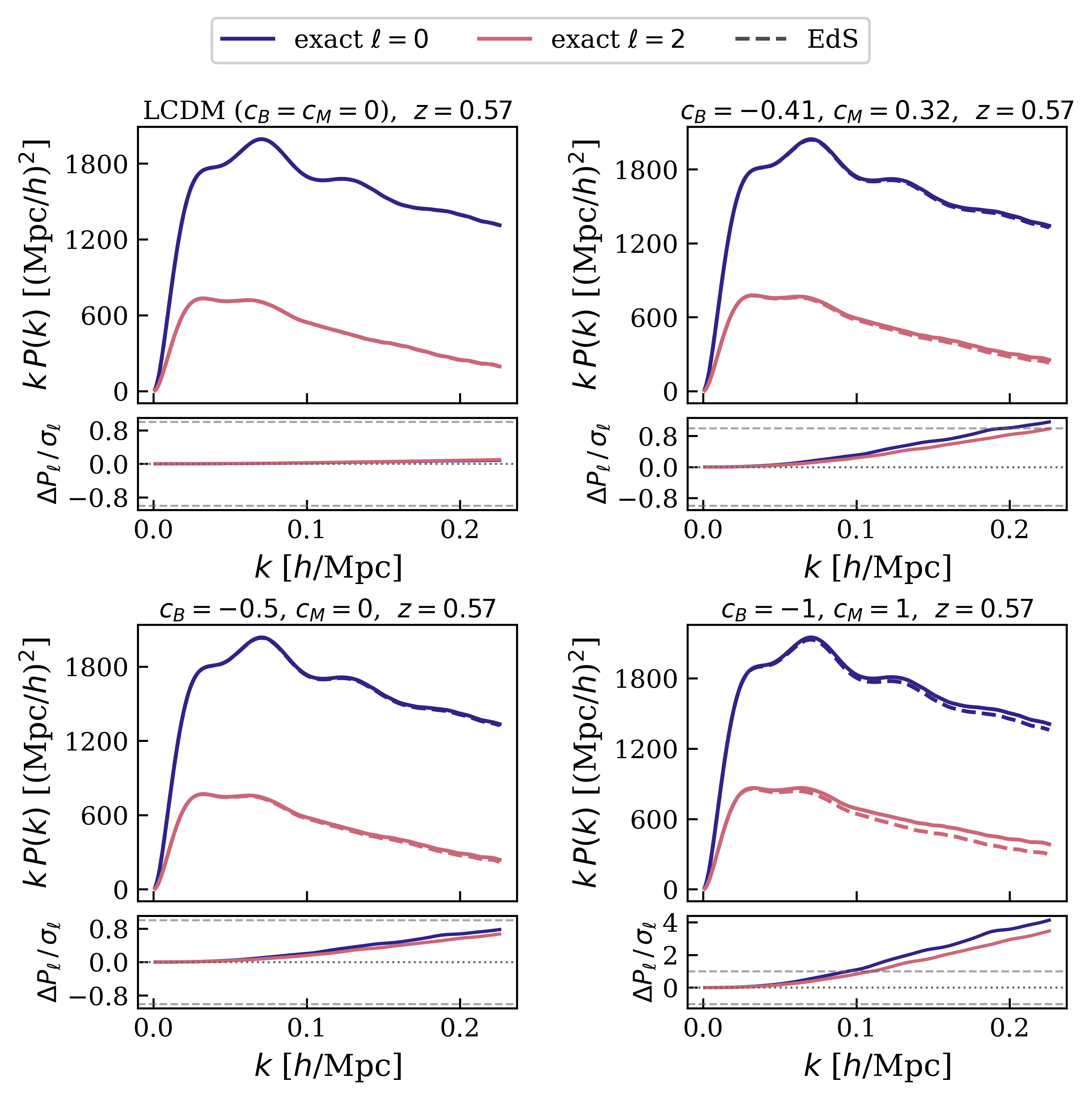}
    \caption{Monopole (indigo) and quadrupole (rose) of the one-loop galaxy power spectrum at $z=0.57$ for several values of the $\Omega_{\rm DE}$ parametrization. Exact time evolution (solid) is compared with the EdS approximation (dashed). EFT parameters are fixed to values fit to BOSS DR12 for the fiducial case $(c_B=-0.24,\,c_M=0.22)$ and held fixed across all panels. The lower panels show the difference weighted by the BOSS covariance diagonal, with horizontal dashed lines marking the $1\sigma$ boundary. The EdS approximation remains well below $1\sigma$ in $\Lambda$CDM; for the fiducial modified gravity case it approaches $1\sigma$ at small scales, and for the extreme case $(c_B=-1,\,c_M=1)$ it reaches $\sim 4\sigma$ near $k_\mathrm{max}$.}
    \label{fig:eds_vs_exact_boss_error}
\end{figure}

The Einstein--de Sitter (EdS) approximation replaces exact time-dependent perturbation-theory kernels with their matter-dominated counterparts and is standard in EFTofLSS analyses. As shown in \cref{fig:eds_vs_exact_boss_error}, this approximation is well justified in $\Lambda$CDM: the difference between the EdS approximation and the exact time-dependent treatment of the one-loop monopole and quadrupole remains well below the $1\sigma$ level (relative to the BOSS DR12 covariance diagonal) over the full range $k \leq 0.23\,h\,\mathrm{Mpc}^{-1}$. Nevertheless, the EdS ansatz remains an approximation, motivated primarily by its computational convenience, and its accuracy need not be assumed to carry over to extended theories of gravity. Rather than validating it separately in this regime, we instead make calculations with the exact time-dependent kernels computationally feasible, removing the need to trade accuracy for efficiency in the first place.

To illustrate why this matters, \cref{fig:eds_vs_exact_boss_error} also compares exact and EdS time evolution for several choices of $\alpha_i \propto \Omega_\mathrm{DE}$, keeping the EFT parameters fixed to the BOSS DR12 best-fit values at $z=0.57$ described in \cref{subsec:parameters_and_priors}; the remaining panels vary the $\alpha$-parameters while holding these fixed, isolating the impact of the kernel time evolution from shifts in the EFT parameters. For the fiducial model, the quadrupole difference approaches $1\sigma$ at $k=0.23\,h\,\mathrm{Mpc}^{-1}$, while for the more extreme case $(c_B=-1,c_M=1)$ it reaches $\sim 4\sigma$. The effect grows toward smaller scales, where EFT counterterms become increasingly relevant, and its magnitude grows with $|c_B|,|c_M|$, i.e.\ non-uniformly across parameter space. A discrepancy of this size and shape could plausibly distort the shape of the $c_B$--$c_M$ posterior, rather than only its best-fit value.

This significance is specific to the BOSS DR12 precision and effective redshift ($z=0.57$). For surveys such as DESI and Euclid, the per-multipole precision $\sigma_\ell$ changes with redshift; furthermore,  for the $\alpha_i \propto \Omega_\mathrm{DE}$ parametrization adopted here, the underlying EdS/exact discrepancy itself change with redshift, since the modification weakens as $\Omega_\mathrm{DE}(z)$ decreases at earlier times. The net effect on significance therefore depends on the interplay between these two trends and cannot be inferred from the BOSS comparison alone, underscoring that the size of the EdS approximation's impact is both survey- and redshift-dependent in a way that no single benchmark can capture. This is precisely why we do not attempt to characterize the EdS error in general: rather than establishing its accuracy for each survey and redshift of interest, we sidestep the question by making the exact time-dependent calculation efficient enough for routine use.

Our direct ODE approach yields exact time-dependent kernels at only modest additional computational cost: for the minimal kernel set required by the one-loop calculation, solving the growth ODE once and reusing the resulting Green's function across target redshifts adds only $\sim 10\,\mathrm{ms}$ to a full one-loop monopole-and-quadrupole evaluation costing $\sim 300$--$400\,\mathrm{ms}$ on a single core (\cref{subsec:validation}), a $\lesssim3\%$ overhead relative to the EdS-only cost. We therefore use exact kernels throughout. While we validate our exact-time one-loop matter power spectrum against $N$-body simulations for this parametrization in \cref{subsec:nbody}, that comparison uses exact-time kernels throughout and does not itself test the EdS-vs-exact difference isolated here; a dedicated $N$-body comparison targeting the EdS approximation specifically would be needed to assess its physical impact for analyses that continue to rely on it.

\section{Conclusion}
\label{sec:conclusion}

In this work, we extended the public EFTofLSS code for biased tracers in redshift space, \texttt{PyBird}, in two main directions. First, we implemented support for scale-independent, Vainshtein-screened luminal Horndeski models, both in the EFTofDE and covariant formalisms, as well as the nDGP model. Second, we replaced the original Green's function implementation with a direct ODE solver for the exact time-dependent functions entering the perturbation kernels. This new approach substantially improves computational efficiency while maintaining excellent agreement with the standard method across all tested models. The gain is particularly significant when evaluating multiple redshifts, since the computational overhead relative to a single-redshift calculation remains small, making the pipeline well suited for likelihood analyses.

We applied this framework to constrain the $\alpha_i \propto a^3$ and $\alpha_i \propto \Omega_{\rm DE}$ parameterizations (with $\alpha_T = 0$) using three data combinations: CMB, CMB+FS, and CMB+BAO+FS. We find that the inclusion of the BOSS full-shape (FS) measurements substantially tightens the allowed region in the $c_M$--$c_B$ parameter space compared with CMB-only constraints. The $a^3$ parameterization generally permits a broader range of $c_M$ values than the $\Omega_{\rm DE}$ case, while projection effects have only a limited impact on the inferred constraints for the models and datasets considered here. Beyond these phenomenological parameterizations, we also presented representative one-loop power spectrum predictions for the cubic Galileon and nDGP models at fixed fiducial parameters, illustrating that the pipeline can be readily applied to covariant luminal Horndeski theories and to modified perturbation sectors beyond the $\alpha_i$ framework. A full Bayesian analysis of these models is deferred to future work.

We also quantified the impact of replacing the exact time-dependent perturbation kernels with their Einstein--de Sitter (EdS) approximation. At fixed EFT parameters, the discrepancy between the EdS and exact predictions for the monopole and quadrupole increases with the departure from $\Lambda$CDM. While it remains well below the BOSS $1\sigma$ uncertainty for $\Lambda$CDM, it reaches several standard deviations for more extreme modified-gravity scenarios within our numerical pipeline. Since this discrepancy depends on the model parameters, it may bias posterior distributions in a way that cannot be inferred from fixed-parameter comparisons alone. A definitive assessment would therefore require repeating the full MCMC analysis using the EdS approximation, which we leave for future work. Because the direct ODE approach provides the exact time-dependent kernels at only a modest additional computational cost, we adopt the exact treatment throughout this work. Although we validated the resulting one-loop matter power spectrum against $N$-body simulations for the phenomenological parameterizations considered here, a dedicated comparison targeting the EdS approximation would ultimately be required to assess its physical accuracy independently of the numerical implementation.

Several natural extensions of this work remain. On the theoretical side, the framework can be generalized beyond the one-loop power spectrum to include higher-order observables such as the bispectrum and the two-loop power spectrum. It can also be expanded to encompass a wider range of modified gravity scenarios, including models with different screening mechanisms. On the observational side, the computational efficiency achieved here makes the pipeline well suited for analyses of forthcoming Stage IV galaxy surveys, where increasingly precise measurements will place stringent demands on theoretical modeling.

Overall, this work provides a systematic, flexible, and computationally efficient framework for extending EFTofLSS analyses beyond $\Lambda$CDM. By combining exact time-dependent perturbation theory with a practical implementation in \texttt{PyBird}, it enables robust one-loop predictions for a broad class of modified gravity theories and paves the way for precision tests of gravity with current and upcoming large-scale structure observations.

\acknowledgments

 We thank Petter Taule for a useful discussion during the initial stage of this project. We thank Himanish Ganjoo for providing us with the N-body data. SV, AS, and DdB acknowledge support from the European Research Council under the H2020 ERC Consolidator Grant “Gravitational Physics from the Universe Large scales Evolution” (Grant No. 101126217 — GraviPULSE). AS acknowledges support from the NWO and the Dutch Ministry of Education, Culture and Science (OCW), through ENW-XL Grant OCENW.XL21.XL21.025 DUSC. GY acknowledges support from the Swiss National Science Foundation.

\newpage

\appendix

\section{Green's Functions, Time-Dependent Functions and Kernels}
\label{sec:app1}
In this section, we present the equations governing the time-dependent functions associated with the Green’s functions, following~\cite{Donath:2020abv,Piga:2022mge}. In our convention, the relevant time-dependent functions are given by
\begin{align}
\mathcal{G}_1^\lambda(a) &= \int^1_0 d\tilde{a} \left[G^\lambda_1(a,\tilde{a})f(\tilde{a}) + G_2^\lambda(a,\tilde{a})\mu_{\Phi,2}(\tilde{a})M_1(\tilde{a})\right]\frac{D^2(\tilde{a})}{D^2(a)} \nonumber \\
\mathcal{G}_2^\lambda(a) &= \int^1_0 d\tilde{a} \, G^\lambda_2(a,\tilde{a})\left[f(\tilde{a}) - \mu_{\Phi,2}(\tilde{a})M_1(\tilde{a})\right]\frac{D^2(\tilde{a})}{D^2(a)} \nonumber \\
\mathcal{U}_1^\lambda(a) &= \int^1_0 d\tilde{a} \left[G_1^\lambda(a,\tilde{a}) \mathcal{G}^\delta_1(\tilde{a})f(\tilde{a}) + G_2^\lambda(a,\tilde{a})M_1(\tilde{a})\left(\mu_{2,\Phi}(\tilde{a})\mathcal{G}^\delta_1(\tilde{a}) + M_2(\tilde{a})\right)\right]\frac{D^3(\tilde{a})}{D^3(a)} \nonumber \\
\mathcal{U}_2^\lambda(a) &= \int^1_0 d\tilde{a} \left[G_1^\lambda(a,\tilde{a}) \mathcal{G}^\delta_2(\tilde{a})f(\tilde{a}) + G_2^\lambda(a,\tilde{a})M_1(\tilde{a})\left(\mu_{2,\Phi}(\tilde{a})\mathcal{G}^\delta_2(\tilde{a}) - M_2(\tilde{a})\right)\right]\frac{D^3(\tilde{a})}{D^3(a)} \nonumber \\
\mathcal{V}^\lambda_{11}(a) &= \int^1_0 d\tilde{a} \left[G_1^\lambda(a,\tilde{a}) \mathcal{G}^\theta_1(\tilde{a})f(\tilde{a}) + G_2^\lambda(a,\tilde{a})M_1(\tilde{a})\left(\mu_{2,\Phi}(\tilde{a})\mathcal{G}^\delta_1(\tilde{a}) + M_2(\tilde{a})\right)\right]\frac{D^3(\tilde{a})}{D^3(a)} \nonumber \\
\mathcal{V}^\lambda_{21}(a) &= \int^1_0 d\tilde{a} \left[G_1^\lambda(a,\tilde{a}) \mathcal{G}^\theta_2(\tilde{a})f(\tilde{a}) + G_2^\lambda(a,\tilde{a})M_1(\tilde{a})\left(\mu_{2,\Phi}(\tilde{a})\mathcal{G}^\delta_2(\tilde{a}) - M_2(\tilde{a})\right)\right]\frac{D^3(\tilde{a})}{D^3(a)} \nonumber \\
\mathcal{V}^\lambda_{12}(a) &= \int^1_0 d\tilde{a} \, G^\lambda_2(a,\tilde{a})\left[\mathcal{G}^\theta_1(\tilde{a}) f(\tilde{a}) - M_1(\tilde{a}) \left( \mu_{\Phi,2}(\tilde{a})\mathcal{G}^\delta_1(\tilde{a}) + M_2(\tilde{a})\right)\right]\frac{D^3(\tilde{a})}{D^3(a)} \nonumber \\
\mathcal{V}^\lambda_{22}(a) &= \int^1_0 d\tilde{a} \, G^\lambda_2(a,\tilde{a})\left[\mathcal{G}^\theta_2(\tilde{a}) f(\tilde{a}) - M_1(\tilde{a}) \left( \mu_{\Phi,2}(\tilde{a})\mathcal{G}^\delta_2(\tilde{a}) - M_2(\tilde{a})\right)\right]\frac{D^3(\tilde{a})}{D^3(a)},
\end{align}
with $M_1$ and $M_2$ defined as
\begin{equation}
M_1(a) = \frac{1}{f(a)}\left(\frac{3\Omega_m(a)}{2}\right)^2, \qquad
M_2(a) = \frac{3}{4}\,\mu_{\Phi,22}(a)\,\Omega_m(a).
\end{equation}

The kernels $K^{(n)}_\lambda$ ($\lambda \in \{\delta,\theta\}$) can be constructed using the time-dependent functions~\cite{Donath:2020abv,Piga:2022mge}:
\begin{align}
K_\lambda^{(1)}(\vec{q}_1,a) &= 1 \nonumber \\
K_\lambda^{(2)}(\vec{q}_1,\vec{q}_2,a) &= \alpha_s(\vec{q}_1,\vec{q}_2)\mathcal{G}_1^\lambda(a) + \beta(\vec{q}_1,\vec{q}_2)\mathcal{G}_2^\lambda(a)\\
K_\lambda^{(3)}(\vec{q}_1,\vec{q}_2,\vec{q}_3,a) &= \sum_{\sigma = 1}^2 \left[\alpha^\sigma(\vec{q}_1,\vec{q}_2,\vec{q}_3)\mathcal{U}^\lambda_{\sigma}(a) + \beta^\sigma(\vec{q}_1,\vec{q}_2,\vec{q}_3) \mathcal{V}^\lambda_{\sigma 2}(a) + \gamma^\sigma(\vec{q}_1,\vec{q}_2,\vec{q}_3)\mathcal{V}^\lambda_{\sigma 1}(a)\right], \nonumber
\end{align}
where
\begin{align}
\alpha_s(\vec{q}_1,\vec{q}_2)
&= \frac{1}{2}\left[\alpha(\vec{q}_1,\vec{q}_2) + \alpha(\vec{q}_2,\vec{q}_1)\right]
&\quad
\alpha^1(\vec{q}_1,\vec{q}_2,\vec{q}_3)
&= \alpha(\vec{q}_3,\vec{q}_{12})\alpha_s(\vec{q}_1,\vec{q}_2)
\nonumber \\
\alpha^2(\vec{q}_1,\vec{q}_2,\vec{q}_3)
&= \alpha(\vec{q}_3,\vec{q}_{12})\beta(\vec{q}_1,\vec{q}_2)
&\quad
\beta^1(\vec{q}_1,\vec{q}_2,\vec{q}_3)
&= 2\beta(\vec{q}_3,\vec{q}_{12})\alpha_s(\vec{q}_1,\vec{q}_2)
\nonumber \\
\beta^2(\vec{q}_1,\vec{q}_2,\vec{q}_3)
&= 2\beta(\vec{q}_3,\vec{q}_{12})\beta(\vec{q}_1,\vec{q}_2)
&\quad
\gamma^1(\vec{q}_1,\vec{q}_2,\vec{q}_3)
&= \alpha(\vec{q}_{12},\vec{q}_3)\alpha_s(\vec{q}_1,\vec{q}_2)
\nonumber \\
\gamma^2(\vec{q}_1,\vec{q}_2,\vec{q}_3)
&= \alpha(\vec{q}_{12},\vec{q}_3)\beta(\vec{q}_1,\vec{q}_2).
\end{align}

\section{Kernels in the Bias Expansion}
\label{sec:app2}

The relevant contributions to the bias expansion kernels are~\cite{Piga:2022mge}:
\begin{align}
K^{(1)}_{\delta_g}(\vec{k},a) &= b_1 \nonumber \\
K^{(2)}_{\delta_g}(\vec{q}_1,\vec{q}_2,a) &= (-b_1 + b_2 + b_4) + b_1 \beta(\vec{q}_1,\vec{q}_2) + \left(b_1 - \frac{2}{7}b_2 \right)\gamma(\vec{q}_1,\vec{q}_2) \nonumber \\
K^{(3)}_{\delta_g}(\vec{q}_1,\vec{q}_2,\vec{q}_3,a) &= \frac{b_1}{3}\beta(\vec{q}_1,\vec{q}_2)\beta(\vec{q}_{12},\vec{q}_3) \nonumber \\
&+ \frac{1}{3}\left(\frac{1}{2}b_1(2\mathcal{G}^\delta_1(a) - 1) + \frac{b_3}{21}\right)\gamma(\vec{q}_1,\vec{q}_2)\beta(\vec{q}_{12},\vec{q}_3) \nonumber \\
&+ \frac{1}{3}\left(\frac{1}{2}b_1(2\mathcal{G}^\delta_1(a) - 1) - \frac{b_3}{21}\right)\gamma(\vec{q}_1,\vec{q}_2)\left(\gamma(\vec{q}_{12},\vec{q}_3) +\alpha_a(\vec{q}_{12},\vec{q}_3)\right) \nonumber \\
&+ \mathrm{cyclic},
\end{align}
where the anti-symmetrized function $\alpha_a$ is defined as
\begin{equation}
\alpha_a(\vec{q}_1,\vec{q}_2) = \frac{1}{2}(\alpha(\vec{q}_1,\vec{q}_2) - \alpha(\vec{q}_2,\vec{q}_1)).
\end{equation}
These bias kernels $K_{\delta_g}^{(n)}$ can be combined with the velocity kernels $K_\theta^{(n)}$ to construct the redshift-space kernels $Z^{(n)} \equiv Z_n$~\cite{Scoccimarro:1999ed,Senatore:2014vja,Piga:2022mge}:
\begin{align}
Z_1(\vec{q}_1,a) &= b_1 + f(a)\mu_1^2 \nonumber \\
Z_2(\vec{q}_1,\vec{q}_2,a) &= K^{(2)}_{\delta_g}(\vec{q}_1,\vec{q}_2,a)
+ f(a)\mu_k^2 K_\theta^{(2)}(\vec{q}_1,\vec{q}_2,a) \nonumber \\
&\quad + f(a)\mu_k k
\Bigg[
\frac{\mu_1}{q_1} K_\theta^{(1)}(\vec{q}_1,a)
\Big(
K^{(1)}_{\delta_g}(\vec{q}_2,a)
+ f(a)\mu_2^2 K_\theta^{(1)}(\vec{q}_2,a)
\Big)
\Bigg] + \mathrm{cyclic} \nonumber \\
Z_3(\vec{q}_1,\vec{q}_2,\vec{q}_3,a) &= K^{(3)}_{\delta_g}(\vec{q}_1,\vec{q}_2,\vec{q}_3,a)
+ f(a)\mu_k^2 K_\theta^{(3)}(\vec{q}_1,\vec{q}_2,\vec{q}_3,a) \nonumber \\
&\quad + f(a)\mu_k k
\Bigg[
\frac{\mu_1}{q_1} K_\theta^{(1)}(\vec{q}_1,a)
\Big(
K^{(2)}_{\delta_g}(\vec{q}_2,\vec{q}_3,a)
+ f(a)\mu_{23}^2 K_\theta^{(2)}(\vec{q}_2,\vec{q}_3,a)
\Big)
\Bigg] \nonumber \\
&\quad + f(a)\mu_k k
\Bigg[
\frac{\mu_{23}}{|\vec{q}_2+\vec{q}_3|}
K_\theta^{(3)}(\vec{q}_2,\vec{q}_3,a)
\Big(
K^{(1)}_{\delta_g}(\vec{q}_1,a)
+ f(a)\mu_1^2 K_\theta^{(1)}(\vec{q}_1,a)
\Big)
\Bigg] \nonumber \\
&\quad + \mu_k^2 k^2 f(a)^2
\frac{\mu_2 \mu_3}{q_2 q_3}
\Bigg[
K_\theta^{(1)}(\vec{q}_2,a)\,
K_\theta^{(1)}(\vec{q}_3,a)
\Bigg.
\nonumber \\
&\qquad \qquad \qquad \qquad \qquad \qquad \times \Bigg.
\Big(
K^{(1)}_{\delta_g}(\vec{q}_1,a)
+ f(a)\mu_1^2 K_\theta^{(1)}(\vec{q}_1,a)
\Big)
\Bigg] \nonumber \\
&\quad + \mathrm{cyclic}.
\end{align}
Here, we adopt the following notation for convenience:
\begin{equation}
\vec{k} = \sum_{i=1}^n \vec{q}_i, \qquad
\mu_k = \hat{k}\cdot \hat{z}, \qquad
\mu_i = \hat{q}_i\cdot \hat{z}, \qquad
\mu_{ij} = \frac{(\vec{q}_i + \vec{q}_j)\cdot \hat{z}}{|\vec{q}_i + \vec{q}_j|}.
\end{equation}

\section{Full posterior constraints}
\label{app:full_post_constraints}
\begin{figure}[H]
    \centering
    \includegraphics[width=\linewidth]{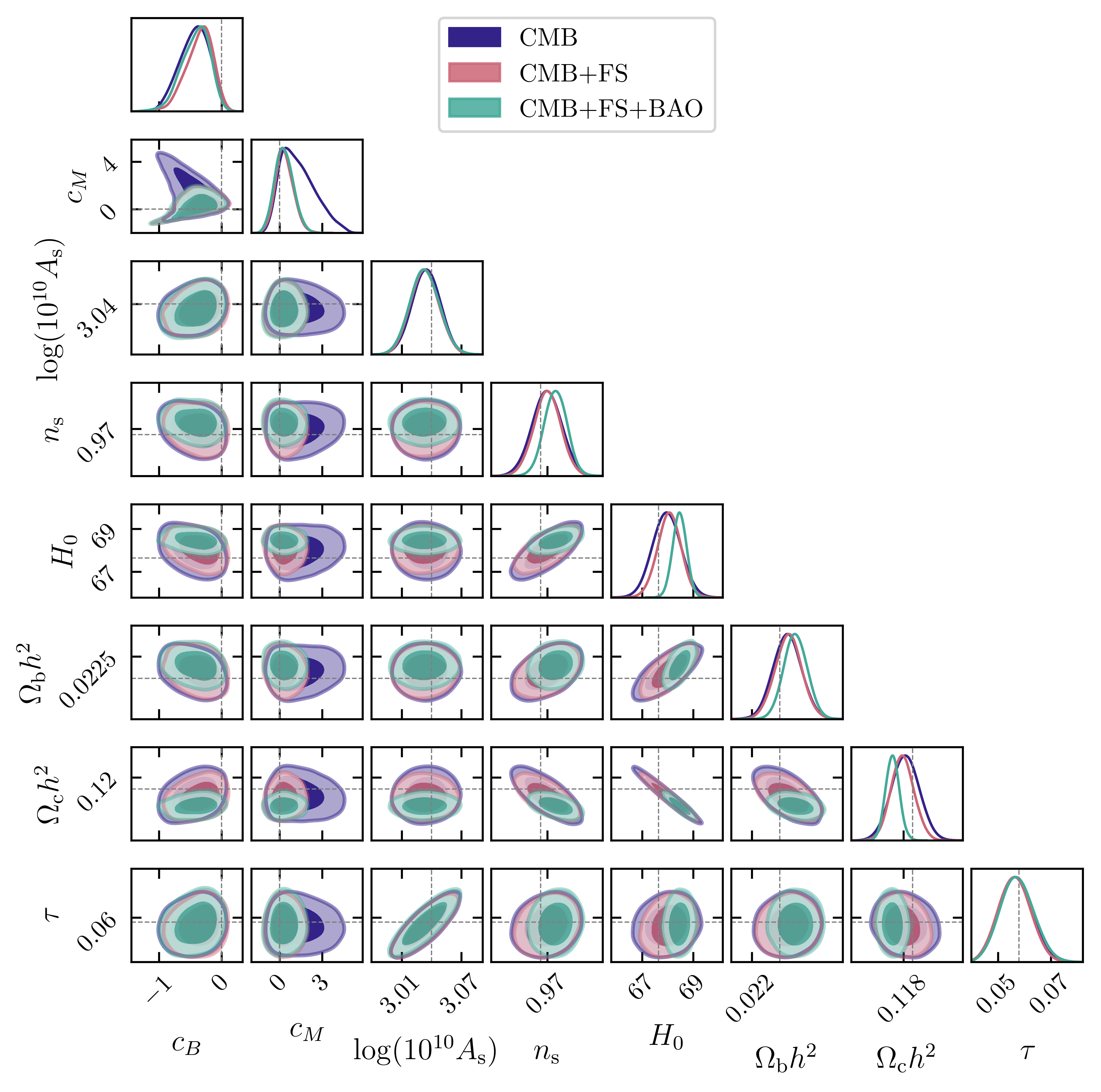}
    \caption{68\% and 95\% CL contours for the EFTofDE amplitude parameters $c_B$ and $c_M$, $\alpha_i=c_i\Omega_\mathrm{DE}$, as well as the six cosmological parameters. Indigo, rose, and teal contours show the CMB-only, CMB+FS, and CMB+FS+BAO constraints, respectively. }
    \label{fig:full_posterior_ode}
\end{figure}

\begin{figure}[H]
    \centering
    \includegraphics[width=\linewidth]{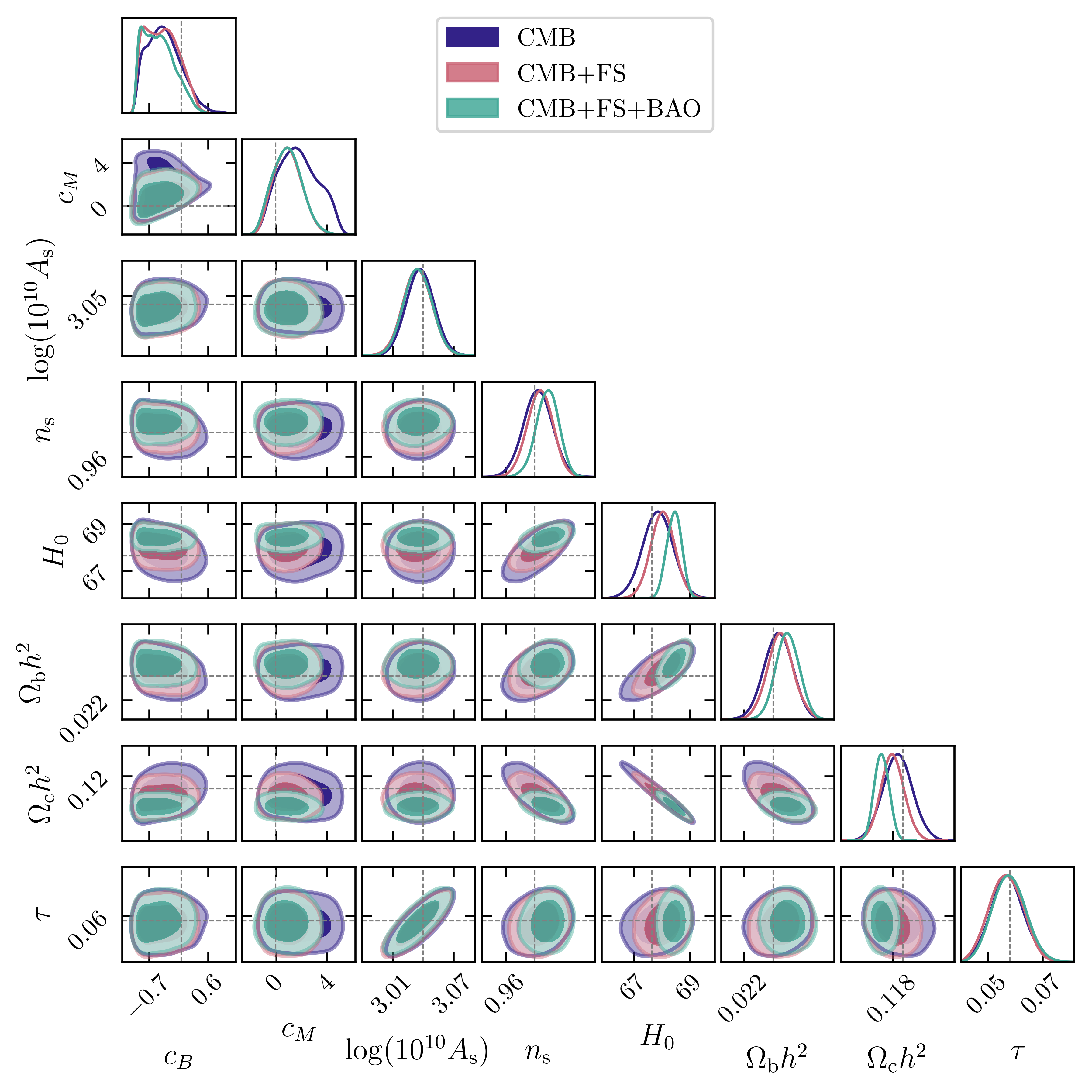}
    \caption{68\% and 95\% CL contours for the EFTofDE amplitude parameters $c_B$ and $c_M$, $\alpha_i=c_i a^3$, as well as the six cosmological parameters. Indigo, rose, and teal contours show the CMB-only, CMB+FS, and CMB+FS+BAO constraints, respectively. }
    \label{fig:full_posterior_cubic}
\end{figure}

\clearpage
\bibliographystyle{JHEP}
\bibliography{references}

\end{document}